\documentclass[5p]{elsarticle}

\usepackage{listings}
\usepackage{color}
\usepackage{soul}

\definecolor{myblue}{rgb}{0,0,0.9}
\definecolor{mygray}{rgb}{0.9,0.9,0.9}
\definecolor{mymauve}{rgb}{0.58,0,0.82}
\usepackage[cmex10]{amsmath}
\usepackage{graphicx}
\usepackage{amssymb}
\usepackage{stmaryrd}

\usepackage{amsthm}
\usepackage{wrapfig}
\usepackage{xspace}
\usepackage{booktabs}
\usepackage{hyperref}

\usepackage[ruled,linesnumbered]{algorithm2e}
\usepackage{algpseudocode}

\usepackage{multirow}
\usepackage{makecell}

\newcommand{\ALOOP}[1]{\ALC@it\algorithmicloop\ #1%
  \begin{ALC@loop}}
\newcommand{\ENDALOOP}{\end{ALC@loop}\ALC@it\algorithmicendloop}

\newcommand{\name}{\texttt{OmClic}\xspace}

\definecolor{garrisonpink1}{rgb}{0.858, 0.188, 0.478}

\journal{Knowledge-Based Systems}

\begin{document}

\begin{frontmatter}

\title{One-to-Multiple Clean-Label Image Camouflage (OmClic) based Backdoor Attack on Deep Learning}

\author[rvt]{Guohong Wang}
\ead{wgh@njust.edu.cn}

\author[ade]{Hua Ma}
\ead{hua.ma@adelaide.edu.au}

\author[dat]{Yansong Gao\corref{cor1}}
\ead{gao.yansong@hotmail.com}

\author[dat]{Alsharif Abuadbba}
\ead{sharif.abuadbba@data61.csiro.au}

\author[wes]{Zhi Zhang}
\ead{zzhangphd@gmail.com}

\author[dat]{Wei Kang}
\ead{wei.kang@data61.csiro.au}

\author[ade]{Said F. Al-Sarawi}
\ead{said.alsarawi@adelaide.edu.au}

\author[rvt]{Gongxuan Zhang}
\ead{gongxuan@njust.edu.cn}

\author[ade]{Derek Abbott}
\ead{derek.abbott@adelaide.edu.au}

\cortext[cor1]{Corresponding author: Yansong Gao}
\address[rvt]{School of Cyber Science and Engineering,
Nanjing University of Science and Technology, Nanjing, JiangSu, China}
\address[ade]{School of Electrical and Electronic Engineering, The University of Adelaide, Australia}
\address[wes]{Department of Computer Science and Software Engineering, University of Western Australia}
\address[dat]{Data61, CSIRO, Australia}

\begin{abstract}
Image camouflage has been utilized to create clean-label poisoned images for implanting backdoor into a DL model. But there exists a crucial limitation that one attack/poisoned image can only fit a single input size of the DL model, which greatly increases its attack budget when attacking multiple commonly adopted input sizes of DL models.

This work proposes to constructively craft an attack image through camouflaging but can fit multiple DL models' input sizes simultaneously, namely \name. Thus, through \name, we are able to always implant a backdoor regardless of which common input size is chosen by the user to train the DL model given the same attack budget (i.e., a fraction of the poisoning rate). With our camouflaging algorithm formulated as a multi-objective optimization, $M=5$ input sizes can be concurrently targeted with one attack image, which artifact is retained to be almost visually imperceptible at the same time. Extensive evaluations validate the proposed \name can reliably succeed in various settings using diverse types of images. Further experiments on \name based backdoor insertion to DL models show that high backdoor performances (i.e., attack success rate and clean data accuracy) are achievable no matter which common input size is randomly chosen by the user to train the model. So that the \name based backdoor attack budget is reduced by $M\times$ compared to the state-of-the-art camouflage based backdoor attack as a baseline. Significantly, the same set of \name based poisonous attack images is transferable to different model architectures for backdoor implant.

\end{abstract}


\end{frontmatter}

\section{Introduction}

The revealed backdoor attacks in 2017~\cite{gu2017badnets,chen2017targeted} on deep learning (DL) models are becoming one of the major barriers of DL trustworthy usage, especially in security-sensitive applications. A backdoored model works normally in the absence of the so-called trigger to be stealthy, but misbehaves once the trigger is present. For example, a backdoored facial recognition model still correctly recognizes Alice as Alice, Bob as Bob if either of them wears a black-framed eye-glass that is the trigger secretly set by the attacker. However, it misclassifies any person who wears this trigger into the Administrator e.g., with higher authorization.
One major attack surface is from the data outsourcing scenario, where a DL model provider/producer outsources the data collection to third parties~\cite{gao2020backdoor}. Data outsourcing is common due to the fact that DL training demands on large amounts of data. However, this requires intensive workforce involved with annotating large datasets or even generating them. Data curation task is thus often outsourced to a third party (e.g., Amazon Mechanical Turk) or volunteers.
In this context, the data could be maliciously poisoned to insert the backdoor once the data is utilized to train a model. 

According to the visual consistency between the image content and its corresponding annotation (i.e., label in classification task), data poisoning based backdoor implant can be divided into two categories: dirty-label poisoning and clean-label poisoning. Generally, the content of the image and its label are different for the dirty-label poisoned image. For example, a dog image stamped with a small trigger is labeled as cat. In contrast, the image content and its label of the clean-label poisoned images are consistent. More details on the dirty-label poisoning and clean-label poisoning can be found in Section~\ref{Section:dirty_clean_poison}.
Almost a majority of existing studies are on the dirty-label poisoning~\cite{gu2017badnets,nguyen2020input,liu2020reflection, liu2017trojaning, wenger2021backdoor,qiu2023towards,ma2023horizontal}. However, the dirty-label poisoning attack is challenging to be survived when the attacker cannot control the model training such as in the model-outsourcing scenario. When the user only needs to outsource the data collection or annotation task, the user will train the model by himself/herself. In that case it is common in real-world, the collected data can undergo human inspection to check whether the image content is consistent with the label. The dirty-labeled images will be rejected in this case. In addition, the reputation of the data provider can be infringed and penalized. 

The clean-label poisonous images retain the labels to be consistent with the images' content. Thus, it can trivially bypass the visual auditing by the data curator. Therefore, the clean-label poisoning poses a realistic security threat to the data collection pipeline even when the curated data undergoes human inspections. However, the clean-label poisoning is less explored due to the stringent label consistency constraint. There are only few works in this research line. Almost all of them build upon the strategy of enforcing the difference between the input space and the latent space/representation, so-called feature collision. to create clean-label poisoned images~\cite{shafahi2018poison,turner2019label,luo2022enhancing,salem2021get}. However, the major limitation of such clean-label attacks is model-dependence. That is the \textit{attacker has to know the model architecture and even weights} used by the victim user's model to determine the latent representation of the adversarial image, which is mandatory during the adversarial image optimization process. This restriction renders the conventional clean-label attack ineffective if the model architecture varies or the weights before the layer of latent representation are changed or the model is trained from scratch~\cite{saha2020hidden}. 

To our knowledge, only the camouflage attack~\cite{xiao2019seeing} based clean-label poisoning is model-agnostic for inserting backdoors into DL models~\cite{quiring2020backdooring,ma2022macab}. The camouflage attack abuses the default image resizing function to create adversary images that are with visually clean labels (detailed in Section~\ref{sec:camoufalgeAtt}). To be effective, the image size fed into the model has to be known to the attacker. We note that this is a reasonable and practical knowledge assumption in real-world. Because the commonly used input size of popular model architectures are few and known to the public. For example, the commonly used input sizes of the ResNet are $224\times 224\times 3$ and $112\times 112 \times 3$. The input sizes of different popular models are summarized in Table \ref{tab:input_sizes_diff_model}.

\begin{table}[htbp]
    \centering
    \caption{Common input sizes of popular DL models.}
    \label{tab:input_sizes_diff_model}
    \begin{tabular}{c | c}
    \toprule
         model & input size \\
    \midrule
         DenseNet~\cite{huang2017densely} & 32, 112, 200, 224\\
         ResNet~\cite{He_2016_CVPR} & 112, 224, 336, 448, 560\\
         VGG~\cite{Simonyan15} & 224, 256, 512\\
         AlexNet~\cite{krizhevsky2017imagenet} & 256, 512\\
         EfficientNet~\cite{tan2019efficientnet} & 224\\
    \bottomrule
    \end{tabular}

\end{table}

A crucial constraint of the data poisoning attack is the poison rate or the attack budget. The attack budget should be as small as possible to be stealthy and efficient to the attacker. For the former, if the poisoning rate is high, it means that the samples of the target class will be notably high, which could be suspicious even for the clean-label attack. For the latter, it means the attacker can spend less effort or time creating poisoned images. In this context, we note that the existing camouflage attack\cite{xiao2019seeing, quiring2020adversarial} can only target a single model input size per attack image, which is inefficient given the fact that there are always several default input sizes of a popular model. To attack all input sizes simultaneously, the poisoning rate has to be increased as a function of the number of targeted input sizes. For example, if a 1\% poison rate can implant a backdoor to the ResNet model given an input size, it requires a 3\% poison rate, $3\times$ higher, to attack three common input sizes concurrently, which consequentially increases the attack budge and becomes less stealthy and efficient.

We address such a crucial limitation by crafting a camouflaged attack image that can target multiple model input sizes simultaneously to fundamentally obviate the requirement of linearly increasing the attacking budget or the poisoning rate upon the existing state-of-the-art (SOTA) camouflage attack~\cite{xiao2019seeing}. The SOTA is incapable of targeting multiple input sizes but a single input size (detailed in Section~\ref{sec:overview}). Consequentially, with the same attack budget, we are able to always implant a backdoor to the model as long as the user adopts any one of the common input sizes to train the models. 

\begin{figure}
    \centering
    \includegraphics[width=0.5\textwidth]{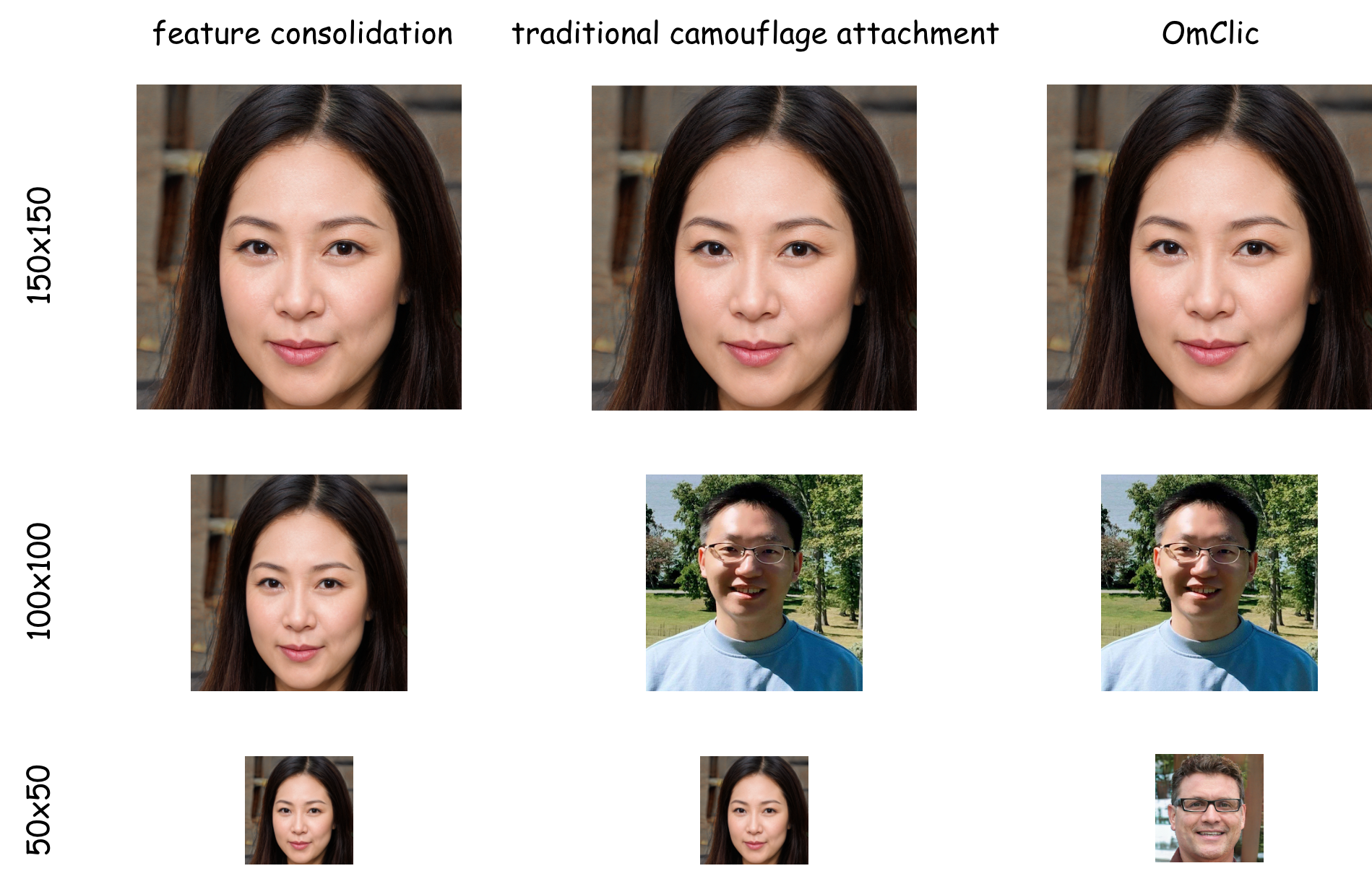}
    \caption{Feature consolidation cannot compromise image semantics of any input size through abusing image resize operation. Traditional camouflage attack can only compromise one input size e.g., $100\times 100$. Our \name now can compromise multiple input sizes e.g., $100\times 100$ and $50\times 50$.}
    \label{fig:intro-diff}
\end{figure}

As shown in Figure~\ref{fig:intro-diff}, we visually compare the resized images of differing sizes when feature consolidation, existing camoufage attack and our \name are launched to abuse the resize operation. Here, feature consolidation, or feature fusion, tackles the inherent challenge of effectively utilizing images with varying resolutions. While low-resolution images provide an initial estimation of an object's spatial location, they may lack the capacity to capture intricate details. On the other hand, high-resolution images offer more nuanced information and encompass richer semantic cues. Throughout the feature fusion process, maintaining semantic consistency across all resolutions remains a key consideration. Note that feature consolidation cannot visually disrupt semantic information through abusing image downsize operation. In traditional camouflage techniques, which exploit the image downsize operation to alter image semantics, an attack image typically affects only one input size of a DL model. This limitation has been demonstrated by Xiao \textit{et al.}~\cite{xiao2019seeing} and subsequent studies~\cite{quiring2020backdooring,ma2022macab}. More specifically, as in Figure~\ref{fig:intro-diff} results in a semantic distinction only at the resolution of $100\times100$, corresponding to its target input size. In contrast, our \name{} introduces a novel approach where a single crafted attack image can simultaneously compromise multiple image input sizes (e.g., $100\times100$ and $50\times50$) of DL models.

Our contributions are summarized as follows:
\begin{itemize}
    \item We propose \name\footnote{It can be pronounced as Oh My Click.}, the first one-to-multiple camouflage attack, that can target multiple input sizes given a single crafted attack image. We formulate the attack image crafting with a multi-objective optimization to automate the attack image generation.

    \item We comprehensively evaluate \name with diverse types of images (i.e., facial images, landscape images) under various settings. Its outstanding performance is affirmed through quantitative and qualitative comparisons with the SOTA.

    \item We demonstrate the practicality of backdoor attacks leveraging the \name through extensive experiments on three datasets: PubFig, STL, and Tiny-ImageNet. The backdoor can always be successfully inserted compared to the baseline backdoor attack regardless of whether any of the targeted model input sizes are chosen by the victim user with the same poisoned set (i.e., only six attack images are sufficient in the facial recognition case study).
\end{itemize}

The rest of the paper is organized as follows. Some necessary background is presented in Section~\ref{sec:relatedwork}. Section~\ref{sec:OmClic} gives an overview of the \name, followed by elaborations on its implementations. Section~\ref{sec:omclicEva} comprehensively and quantitatively evaluates \name on diverse type of images with various setting considerations, as well as comparisons with the SOTA~\cite{xiao2019seeing}. Backdoor attacks based on \name are presented and extensively evaluated in Section~\ref{sec:backdoor}. We discuss \name enabled backdoor attacks further in Section~\ref{sec:disc}, especially with providing an easy-to-deploy lightweight prevention method to mitigate the \name. Section~\ref{sec:conclusion} concludes this work.

\section{Related Work}\label{sec:relatedwork}

\subsection{Backdoor Attack Scenario}
A backdoored model behaves normally for inputs without the trigger but misbehaves as the attacker-specified once the attacker presents his/her secretly chosen trigger in the input~\cite{gao2020backdoor}. For example, supposing the trigger is a sun-glass, any person, e.g., person A, without wearing it will still be recognized as person A by the backdoored facial recognition model. However, he/she will be recognized as the administrator by the backdoored model once the sun-glass is worn. There are a number of real-world scenarios that can introduce backdoor into the DL model as long as the model or its training dataset can be tampered by the attacker. These means are model outsourcing~\cite{gu2017badnets}, dataset outsourcing~\cite{shafahi2018poison}, distributed machine learning~\cite{bagdasaryan2020backdoor}, pretrained model reusing~\cite{yao2019latent}, or even through vulnerable code called by the DL framework~\cite{bagdasaryan2021blind}, and fault injection after model deployment~\cite{qi2022towards}.

\subsection{Data Poisoning based Backdoor}\label{Section:dirty_clean_poison}
Data outsourcing is one of the three most common scenarios (i.e., the first three) compared to the last three scenarios. Due to the hardness of collecting some specific data (i.e., medical) or the intensive involved labor, it is common that a model trainer outsources the data collection or/and data annotation to third parties. For instance, the Amazon Mechanical Turk\footnote{\url{https://www.mturk.com/}} is such as platform where one can issue dataset outsource tasks. The annotation of a commonly used FLIC dataset~\cite{modec13} for object detection task was outsourced to Amazon Mechanical Turk.
In addition, some data collections rely on volunteer contributions. Moreover, some large-scale datasets, e.g., ImageNet~\cite{deng2009imagenet} are crawled from the Internet and annotated through crowdsourcing~\cite{deng2009imagenet}. For all these cases, the data can be tampered with before being received by the data curator. A small fraction (i.e., 0.06\%~\cite{gao2021design}) of tampered or poisoned data can essentially succeed in inserting a backdoor into a DL model trained upon it.

Data poisoning can be generally divided into two categories: dirty-label poisoning and clean-label poisoning. Gupta et al.~ carried out two additional attacks in addition to the targeted attack: a random label flipping attack•and a random input data poisoning attack. The former is dirty-label poisoning, and the latter is clean-label poisoning.
The difference between these two poisoning categories is following:
\begin{itemize}
    \item Dirty-label poisoning. The labeling of samples is inconsistent with the semantics of these samples. which is trivially achievable by simply altering the label of a poisoned sample that contains the trigger. This is not stealthy for human inspection.
    \item Clean-label poisoning. It ensures the consistency between the poisoned image content and its annotated label. 
    Thus, human inspector can not find any irregularity attributing the consistency.
\end{itemize}

To craft clean-label poisonous images, the majority of studies~\cite{shafahi2018poison,turner2019label,luo2022enhancing,salem2021get} utilize the feature collision attack. For example, a poisoned face image of person A is labeled as person A, which is visually unsuspicious due to the consistency between the image content that is the input space or pixel space and the annotation. However, when it is fed into a DL model, its latent representation (i.e., from the first fully connected layer of a CNN model) in latent space is in fact equal to person B. This can be exploited to perform backdoor attacks through clean-label data poisoning~\cite{saha2020hidden}. That is, the feature of poisoned image A collides with image B in the latent space even though they are different in the input space. Generally, the perturbed/poisoned A's image feature representation is similar to any other person's face image (i.e., person B) \textit{stamped with a trigger} (i.e., sun-glass). The model trains on the poisoned dataset and learns a backdoor/association between the trigger and targeted person A, thus misbehaving backdoor effect to misclassify any person with the trigger to person A.

However, clean-label poisoning upon feature collision has a crucial limitation that a feature extractor to extract the latent representation should be known by the attacker. This means the attacker often needs to have white-box knowledge of the feature extractor (i.e., the victim model). Generally, poisonous image crafting in this context is (victim) model dependent.

\subsection{Camouflage Attack}\label{sec:camoufalgeAtt}
The other means of crafting clean-label poisonous images is through the camouflage attack\cite{xiao2019seeing} by abusing the default resizing operation provided by commercial DL frameworks~\cite{quiring2020backdooring,ma2022macab, chen2022white}. In \cite{chen2022white}, Chen et al. extended camouflage attacks by utilizing five types of pre-processing modules common in DL systems. For the camouflage attacked image, its visualization is different before and after resizing operation. Note that the image size (i.e., the resolution up to $4032\times3024$ for images taken by iPhone 13) is always larger than the input size of a given DL model (see Table~\ref{tab:input_sizes_diff_model}). These large images will be downsized into the model's acceptable input size by calling the default resizing function before feeding them into the model for either training or inference. Therefore, an attacker can create an attack image (i.e., person A's face image) seen by the data curator that will become the target image (i.e., person B/C's face image with a trigger) seen by the model. Here, the attack image retains its consistency between the image content and the annotation. Obviously, once a DL model trains on these poisoned images, it will be backdoored. So that it will classify any person with the trigger to person A who is the attacker-targeted person such as the administrator.

Despite this clean-label poisoning attack exhibiting a main merit of being independent on DL models, it is dependent on the model input size targeted. For example, if the targeted size is $224\times 224 \times 3$, its effect will not function if the model user chooses any other input size e.g., the other common option of $112\times 112 \times 3$ (see Table~\ref{tab:input_sizes_diff_model}). When performing backdoor attacks, the attacker has to linearly increase its poison rate (i.e., using more poisonous images) if the attacker targets multiple model input sizes. This is undesirable as it is less stealthy and increases the attack budget. In the following, we present \name that can cover multiple model input sizes given the same poisonous image without increasing the poisoning rate at all.

Almost all existing image-resizing attacks predominantly rely on optimization formulations to generate attack images. In contrast, our approach employs a multi-objective optimization strategy, deviating from the single-objective optimization employed in previous image-resizing attacks. This multi-objective optimization facilitates the concealment of multiple target images within the same source image, enabling \name.

The way to craft an attack image might utilize direct computation upon reversing the interpolation algorithm or attack image generation through a generative adversarial network. For example, this work~\cite{liu2019query} opts for the former means, however, the produced attack image is often not semantically good, which requires some trials. Despite the possibility of using GAN, there has been no such work done for image resizing attacks. As a future work, it is interesting to explore the feasibility.

\section{One-to-Multiple Clean Label Image Camouflage}\label{sec:OmClic}


\subsection{Overview}\label{sec:overview}

\begin{figure}[ht]
    \centering
    \includegraphics[width=\linewidth]{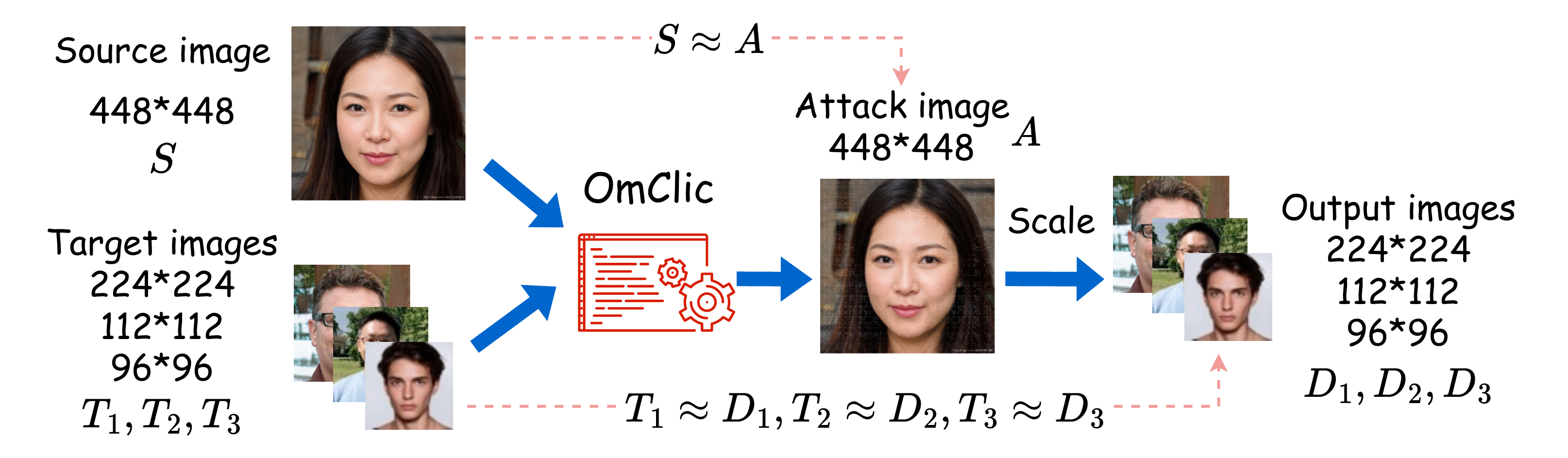}
    \caption{\name overview. Three target images with different semantic contents and sizes are used for example.}
    \label{fig:overview}
\end{figure}

The overview of the One-to-Multiple Clean Label Image Camouflage (\name) is shown in Figure \ref{fig:overview}. The aim is to disguise multiple target images (i.e., $k$ $T$s) in the same source image ($S$)---$k=3$ in the example. The manipulated source image $S$ is the attack image $A$ that will be received by the victim user who uses it to train a DL model. The attack image $A$ is visually close to the source image---its annotation (i.e., label) is consistent with its content (i.e., the lady is labeled with an correct name). However, once it is used to train a DL model, its content becomes semantically similar to the target image $T$ due to the abuse of the default \textsf{scale} function provided by mainstream DL frameworks. More precisely, $D_1 \approx T_1$ where $D_1=$\textsf{scale$_1$}($A$). By stamping a trigger on a fraction of different target images $T$s before disguising each into an attack image $A$, a backdoor will be inserted into the downstream DL models, as experimentally evaluated in Section~\ref{sec:backdoor}. 

In this context, the key to \name is to strategically craft the attack image. The \name aim is to disguise $k$ target images rather than a single target image into the source image as performed by Xiao \textit{et al.}, the SOTA~\cite{xiao2019seeing}. The $k$ target images can have different semantic contents (i.e., faces of different persons or faces of the same person but at different angles), or different image sizes (i.e., the face of the same person at the same shooting setting but different resolution/size), or a combination of above two scenarios, as exemplified in Figure~\ref{fig:overview}.

\vspace{0.2cm}
\noindent{\bf Challenges and Our Solution.} Intuitively, the methodology devised by Xiao \textit{et al.}~\cite{xiao2019seeing}, exchangeably referred to as SOTA, can be consecutively applied to each of these $k$ target images, hopefully, to gain an attack image retaining the deceive effect. However, our trials showed this is not immediately applicable. Firstly, the disguising operation tends to often fail due to the non-existence of optimization solution under the relatively too strong constraints set by the SOTA. Generally, this is because the SOTA transforms the attack into a convex optimization problem. Once the constraints (i.e., the perturbation amplitude on the attack image and the difference between the output image resized from the attack image and the target image) are enforced, it might not always converge to a satisfactory solution, thus causing a failure. 
Secondly, the SOTA camouflage is extremely computationally heavy, which renders unbearable time overhead, especially for relatively large-size attack images, even when camouflaging merely a single target image. Generally, this is because the SOTA solves the pixel perturbation in a fine-grained manner, e.g., line by line of the image. This inevitably invokes the convex-concave programming toolkit much more frequently, rendering costly computation (i.e., the overhead is dependent on the image size).

The \name resolves the above shortcomings through two major means. 
Firstly, we transform the \name camouflage attack into a distinct multi-objective optimization problem~\cite{deb2014multi}. This overcomes frequent failure of the SOTA during the optimization process. Note that the multi-objective optimization naturally fits our one-to-multiple attack, since multiple target images have to be disguised simultaneously. Secondly, we solve the pixel perturbation per channel (i.e., a colorful image has three channels). Therefore, the number of invocations of the optimization toolkit is independent of the image size, and importantly, extremely less (i.e., only three invocations are required for a colorful image). Consequentially, the computation load of \name is very efficient.

\subsection{Implementation}\label{omclic-imple}

We first define some notations. The $m$ and $n$, respectively, denote the number of rows and columns of source image size, and $c$ denotes the number of channels---in particularly, $c=3$ for colorful images. Similarly, $m_j$ and $n_j$ denote the $j_{\rm th} \in \{1,...,k\}$ image size of the $j_{\rm th}$ target image. Note in the camouflage attack, the target image size is usually smaller than that of the source image. This is aligned with the fact that image downscaling is more common when training the DL model. $a$ denotes the pixel value, which should be in the range of [0,255]. $L_j$ and $R_j$, respectively, denote the left and right constant coefficient matrix when a target image is resized, see the Eq \ref{eq:scale_function}. Note that $L_j$ and $R_j$ are deterministic once the $m$, $n$, $m_j$, and $n_j$ are given---they are known in camouflage attack.

Our primary objective is to ascertain the minimum value of $\Delta$ that ensures the visual resemblance of the attack image to the source image. To address this challenge, we employ a multi-objective optimization approach, akin to seeking a solution within specified constraints. To maintain the semantics of the image and evade human inspection, we need to constrain the difference between the source image and the attack image, which are usually quantified through the so-called $L_n$ norm---$n$ is usually 1, 2 and $\infty$. Here, $L_1$ computes the absolute difference sum of all elements (e.g., pixels given an image), $L_2$ computes the square root value of the sum of each pixel difference across the whole image, and $L_\infty$ describes the largest pixel difference among all pixel elements. Compared to $L_1$ and $L_\infty$, $L_2$ relaxes the space in which to find the better perturbation, and this difference is not sensitive to semantic. Thus, we chose the Euclidean norm as a constraint.
In this context, the relationship between $A$ and $S$ is formalized to be: 
\begin{equation}
    \begin{aligned}
        &A_{m \times n} = S_{m \times n} + \Delta\\
        &\textsf{Obj}: \min(\lVert \Delta \rVert_2)
    \end{aligned}
\end{equation}

To solve $\Delta$, we further formalize the scaling process. Since the scaling size (i.e., the output image size) is fixed, the \textsf{Scale} operation can be expressed as:
\begin{equation}
    \label{eq:scale_function}
    \textsf{Scale}_j(A_{m\times n}) = L_{m_j \times m}*A_{m\times n}*R_{n \times n_j} = T_{m_j \times n_j}, 
\end{equation}
where $j$ is for the $j_{\rm th}$ target image. $L_{m_j \times m}$ and $R_{n \times n_j}$ are two coeffecient matrices that can be stably solved~\cite{xiao2019seeing} given the known scaling size.

Once the scaling size is fixed, the scaling coefficient is stable. From Xiao \textit{et al}~\cite{xiao2019seeing}, this coefficient can be inferred from input and output pairs. For example, the input can be the source image while the output can be the target image or vice versus. In other words, the image content is not matter, the input and output sizes matter.

First of all, we can build the relationship between input and output pairs:
\begin{equation}
    \label{eq:coefficient_1}
    \begin{aligned}
        &L_{m' \times m} * (I_{m \times m} * IN_{max}) = L_{m' \times m} * IN_{max} \\
        &(I_{n \times n} * IN_{max}) * R_{n \times n'} = R_{n \times n'} * IN_{max},
    \end{aligned}
\end{equation}
where $I_{m\times m}$ and $I_{n \times n}$ are both identity matrices. And $IN_{max}$ stands for the max element in the source image (i.e., it can be any scalar excepting $0$ and $1$).

For example, by setting $S = I_{m\times m}*IN_{max}$ and scaling it into an $m' \times m$ image $D_{m' \times m}$, we can infer $L_{m' \times m}$ since:
\begin{equation}
    \label{eq:coefficient_2}
    \begin{aligned}
        &D = \textsf{Scale}(S) = \texttt{unsigned int}(L_{m' \times m} * IN_{max})\\
        &\rightarrow L_{m' \times m(appr)} \approx D / IN_{max}
    \end{aligned}
\end{equation}

Since division is a finite decimal, Eq. \ref{eq:coefficient_2} brings a slight precision loss. To ensure that the sum of elements in each row of the coefficient matrix is one, normalization is applied per row in the coefficient matrix to make it accurate. 
\begin{equation}
    \label{eq:coefficient_3}
    \begin{aligned}
        &L_{m' \times m(appr)}[i,:] = \frac{L_{m' \times m(appr)}[i,:]}{\sum_{j=0}^{m-1}(L_{m' \times m(appr)}[i,j])} \\
        &(i = 0, 1, \cdots, m'-1)
    \end{aligned}
\end{equation}

Eventually, the optimization formula is following:
\begin{equation}
\label{eq:getPertubation}
    \begin{aligned}
        \textsf{Suppose:} \Delta &= A_{m\times n} - S_{m\times n}\\
        \epsilon_j &= \lVert \textsf{Scale}_j(A_{m\times n}) - T_j\rVert_2\quad j = 1, 2, \cdots, k\\
        \textsf{loss} &= \lVert\Delta\rVert_2 + \epsilon_1 + \cdots + \epsilon_k\\
        \textsf{Object:} &\quad \min(\textsf{loss})\\
        \textsf{Constrain:} &\quad \forall a \in A, 0\le a\le 255
    \end{aligned}
\end{equation}

In the equation above, it firstly defines a set of distances(object) (e.g., $\epsilon$) that aim to minimize, e.g., the difference between the attack image and the source image. Afterward, to improve computational efficiency, we employ the weighted sum to transform this multiple-objective optimization into a single-objective optimization (in particular, the \textsf{loss}), assigning a weight of 1. To this end, this problem have been converted to solving the optimization formula loss under a constrain.

\begin{algorithm}
\caption{Generating an attack image with multiple target images.} 
\label{alg:omclic}
    \KwIn{Source: $S\in \mathbb{N}_{m\times n \times c}$;\newline
    Target: $T_1\in \mathbb{N}_{m_1\times n_1 \times c},\cdots, T_k \in \mathbb{N}_{m_k\times n_k \times c}$; \newline
    Scale functions: $\textsf{Scale}_1(), \cdots , \textsf{Scale}_k()$}
    \KwOut{Attack image: $A = S + \Delta \in \mathbb{N}_{m \times n \times c}$}
    $A, \Delta = 0_{m \times n \times c}, 0_{m \times n \times c}$\;
    \For{$i=0$ to $c-1$}{
        Initialize $obj=0$\;
        $A_i = A[:,:,i], \Delta_i = \Delta[:,:,i]$\;
        \For{$j=1$ to $k$}{
            $T_{j_i} = T_j[:,:,i]$\;
            $L_j, R_j = \textsf{GetCoefficient}(m, n, m_j, n_j)$\;
            $obj += \lVert L_j * A_i * R_j - T_{j_i}\rVert_2$
        }
        $obj += \lVert \Delta_i\rVert_2$\;
        $\Delta_i = \textsf{GetPerturbation}(obj, \Delta_i)$\;
        $A[:,:,i] = S[:,:,i]+\Delta_i$
    }
    \Return{$A_{m \times n \times c}$}   
\end{algorithm}

The algorithm implements of \autoref{eq:getPertubation} has been reformulated into the following structure. Lines $2-13$ are to find the optimal perturbation iterating over all three color channels of the image, Lines $5-9$ restricts distances between the source image and target images. Subsequently, line $10$ puts Euclidean norm $L_2$ of perturbation. Then, the optimization procedure to ascertain the optimal perturbation, denoted as \textsf{GetPerturbation()}, is executed through the utilization of an available framework \texttt{cvxpy}. This algorithm returns the attack image at the end.

\section{\name Evaluation}\label{sec:omclicEva}

This section evaluates the \name under varying settings e.g., when different target images with different sizes are embedded into a single source image. 
In addition, we quantitatively compare \name with Xiao \textit{et al.}~\cite{xiao2019seeing} in terms of deceive effect of the attack image and the time overhead. 
Diverse types of images\footnote{These images are from \url{https://wallpaperaccess.com}.} including facial images, animal images, and landscape images, are utilized to comprehensively evaluate the \name. 
Experimental evaluations of \name enabled model agnostic backdoor attacks are deferred to Section~\ref{sec:backdoor}.

\subsection{Different Target Images with Different Output Sizes}

\begin{figure}[h!]
    \centering
    \includegraphics[width=0.47\textwidth]{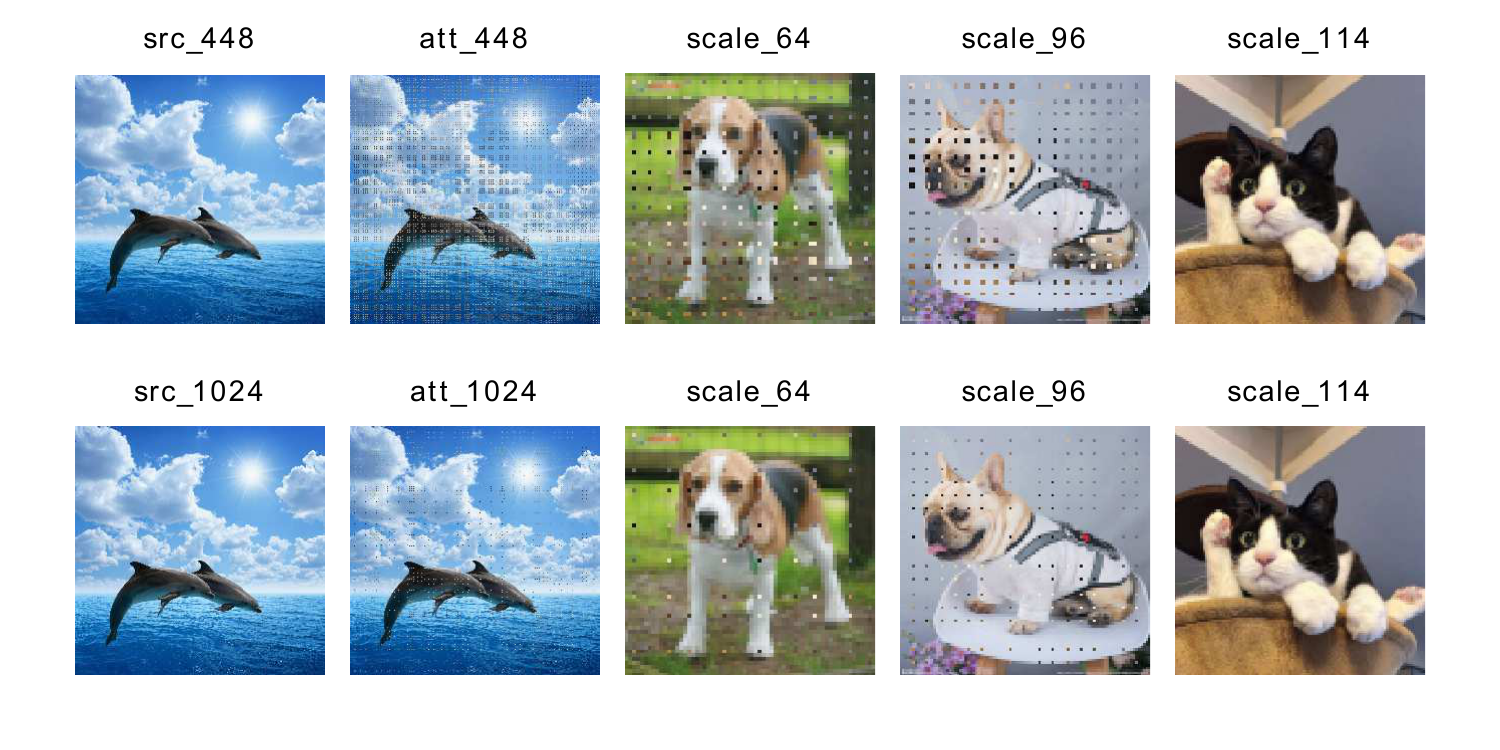}
    \caption{Different target images with different sizes. Animal images are used.}
    \label{fig:diff_tar}
\end{figure}

As shown in Figure \ref{fig:diff_tar}, we embedded three different target images (e.g., dog or cat) into one source image. Three different sizes are $64\times64\times3$, $96\times96\times3$ and $114\times114\times3$, respectively. The source image has a size of $448\times 448 \times 3$ and a larger size of $1024\times 1024 \times 3$ has also been evaluated. In this experiment, the \textsf{scale} function is set to be \texttt{NEAREST}.

Firstly, the attack image (i.e., second column) is similar to the source image (i.e., first column). Secondly, the output images (i.e., third to fifth columns) are visually similar to their corresponding target images.
In addition, we note that when the size of the source image is small, $448\times448\times3$, there are some perceptible artifacts to the output image (i.e., the scaled dog image with a size of $96\times 96\times 3$).
The artifact can be mitigated when the size of the source image increases e.g., $1024\times1024\times3$ in the second row. This means the difference between the target image and the output image becomes small.

The reason is that the performance of the image scaling attack is dependent on the ratio of the source image and the target image. Generally, the target image is inserted into the source image by dispersing delicate noises. It is easier to do this imperceptibly conditioned on a larger ratio between the source image and the target image, because there will be more flexible space allowing these delicate noises to be injected. In the other way around, the higher this ratio, the better (visual) similarity between the attack image and the source image.
As theoretically analyzed by Quiring \textit{et al.}~\cite{quiring2020adversarial}, not all pixels in the source image equally contribute to its scaled version. Only those pixels close to the center of the kernel weigh high, whereas all remaining pixels play a limited role during scaling. This imbalanced influence of the source pixels provides a perfect ground for image-scaling attacks. The adversary only needs to modify those pixels with high weights to control the scaling and can leave the rest of the image untouched. Thus, if the ratio of the source image and the target image is higher, it is easier to find those pixels with higher weights. Consequentially, the similarity between the attack image and the source image will be high.
So there is nearly no notable perturbation on the three output images scaled from the attack image with the size of $1024 \times 1024 \times 3$.

\begin{figure}[ht]
    \centering
    \includegraphics[width=0.47\textwidth]{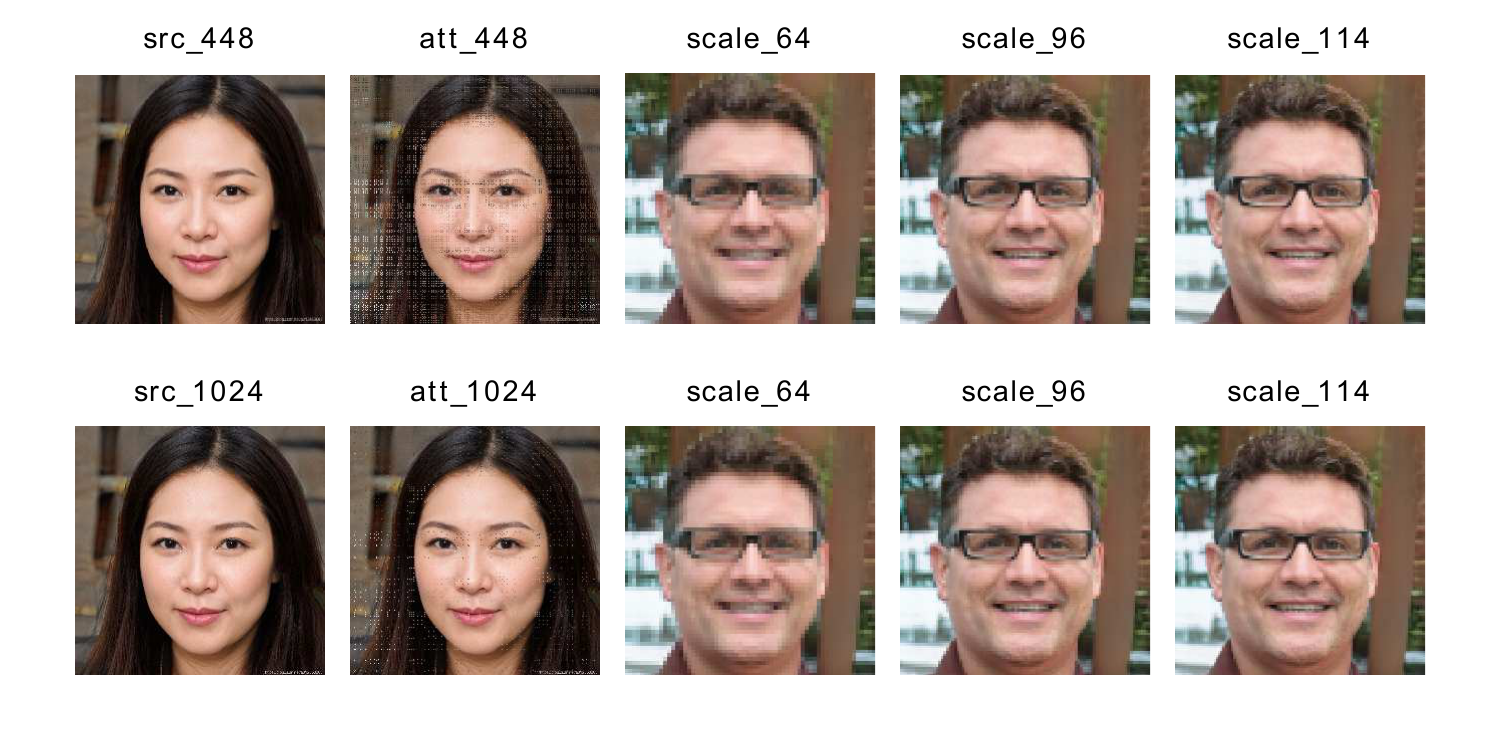}
    \caption{Same target image with different sizes. Face images are used.}
    \label{fig:diff_size}
\end{figure}

\subsection{Same Target Image with Different Output Sizes}
\label{section:multi-size}
Here, we implant three visually same target images but with different sizes $64\times 64 \times 3$, $96\times 96 \times 3$, $114 \times 114 \times 3$ into one source image, which formed attack images (i.e., second column) are shown in Figure \ref{fig:diff_size}. The target images are the same person's face images but with different resolutions in this example.

Even though a small size source image $448\times448\times3 $ is used, there are nearly no perceptible artifacts on these three output images (i.e., columns 3, 4, and 5). There are two potential reasons. Firstly, all target images are visually the same. Secondly, the target images and the source image are all face images, which similarity is also higher than that in Figure~\ref{fig:diff_tar}, where the source image (i.e., dolphin) is quite distinct from those target images (i.e., dog or cat). 

This implies that semantic similarity between the source image and target image, or/and similarity among targets image can exhibit a better \name deceive effect.

In addition, note that a larger source image size is beneficial to the removal of the artifacts brought to the attack image. More precisely, when looking closer (zooming in), the artifacts in the $448\times 448 \times 3$ attack image are perceptible but eliminated when $1024\times 1024 \times 3$ source image is utilized.

\subsection{Same Target Image with Different Resize Functions}\label{sec:diffResize}
\begin{figure}
    \centering
    \includegraphics[width=\linewidth]{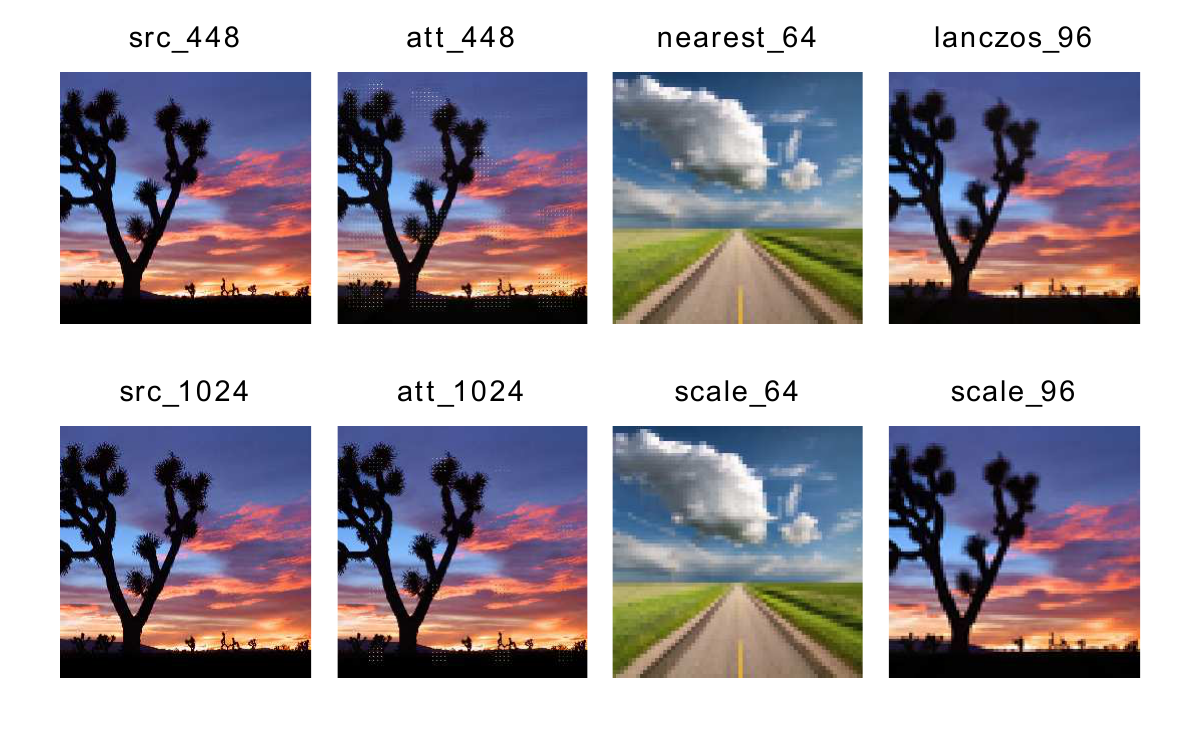}
    \caption{Same target image with different resize functions. Landscape images are used.}
    \label{fig:diff_func}
\end{figure}

In Figure \ref{fig:diff_func}, we set up the case when the same target image is resized by different resize functions into different output sizes. 
During the attack image crafting, the \textsf{scale} function \texttt{NEAREST} is used to disguise the same target image with different sizes $64\times64\times3$ (i.e., third column) and $96\times96\times3$ (i.e., fourth column) into the source image. 

On one hand, if a different resizing algorithm e.g., \texttt{LANCZOS}, is chosen to rescale the attack image to gain the output image e.g., $96\times96\times3 $ in the third column, the output image is semantically similar to the source but not the target image intended by the attacker.
On the other hand, if the attack image is resized to the output image of a $64\times64\times3 $ size with the same algorithm of \texttt{NEAREST}, the output image as expected is nearly the same as the target image. We have evaluated other combinations, e.g., \texttt{NEAREST} is used during attack image crafting and a different \texttt{LANCZOS} function is used to resize the attack image. We found that the camouflage effect can work only when both resize functions are the same during the attack image creation and attack image resize in all our experiments.

\begin{table*}[htb]
    \centering
    \caption{Quantitative similarity comparison between Xiao \textit{\textit{et al.}}~\cite{xiao2019seeing} and \name.}
    \resizebox{\linewidth}{!}{
    \begin{tabular}{ c  c  c  c  c  c  c  c  c  c  c  c  c  c  c  c  c  c }
    \toprule
    \multicolumn{2}{c}{\multirow{4}*{Types}} & \multicolumn{4}{c}{SSIM} & \multicolumn{4}{c}{MSSSIM} & \multicolumn{4}{c}{UQI} & \multicolumn{4}{c}{PSNR} \\
    \cmidrule{3-18} \multicolumn{2}{c}{} & \begin{tabular}{@{}c@{}} Xiao \textit{et al.} \\ \cite{xiao2019seeing} \end{tabular}  & \multicolumn{3}{c}{{\begin{tabular}{@{}c@{}} Ours \\ \name \end{tabular}}} & \begin{tabular}{@{}c@{}} Xiao \textit{et al.} \\ \cite{xiao2019seeing} \end{tabular} & \multicolumn{3}{c}{\begin{tabular}{@{}c@{}} Ours \\ \name \end{tabular}} & \begin{tabular}{@{}c@{}} Xiao \textit{et al.} \\ \cite{xiao2019seeing} \end{tabular} & \multicolumn{3}{c}{{\begin{tabular}{@{}c@{}} Ours \\ \name \end{tabular}} } &\begin{tabular}{@{}c@{}} Xiao \textit{et al.} \\ \cite{xiao2019seeing} \end{tabular} & \multicolumn{3}{c}{\begin{tabular}{@{}c@{}} Ours \\ \name \end{tabular}} \\
    \cmidrule{3-18} \multicolumn{2}{c}{} &
    1 & 1 & 2 & 3 & 1 & 1 & 2 & 3 & 1 & 1 & 2 & 3 & 1 & 1 & 2 & 3 \\
    \midrule
    \multirow{2}{*}{Face} & 448 & 0.744 & 0.742 & 0.565 & 0.472 & 0.942 & 0.942 & 0.887 & 0.844 & 0.90 & 0.90 & 0.833 & 0.79 & 27.469 & 27.483 & 22.136 & 19.422\\
    \cmidrule{2-18} & 1024 & 0.889 & 0.905 & 0.755 & 0.662 & 0.979 & 0.982 & 0.949 & 0.917 & 0.997 & 0.975 & 0.929 & 0.888 & 33.412 & 34.447 & 29.307 & 26.415\\
    \midrule
    \multirow{2}{*}{Animal} & 448 & 0.655 & 0.660 & 0.47 & 0.38 & 0.936 & 0.936 & 0.873 & 0.821 & 0.971 & 0.971 & 0.946 & 0.925 & 25.102 & 25.262 & 19.819 & 17.096\\
    \cmidrule{2-18} & 1024 & 0.881 & 0.865 & 0.665 & 0.567 & 0.982 & 0.980 & 0.943 & 0.907 & 0.994 & 0.992 & 0.979 & 0.966 & 33.518 & 32.113 & 26.977 & 24.079\\ 
    \midrule
    \multirow{2}{*}{Landscape} & 448 & 0.734 & 0.726 & 0.564 &0.474 & 0.944 & 0.942 & 0.892 & 0.847 & 0.839 & 0.838 & 0.801 & 0.778 & 26.574 & 26.413 & 21.262 & 18.547\\
    \cmidrule{2-18} & 1024 & 0.917 & 0.889 & 0.722 & 0.632 & 0.987 & 0.979 & 0.942 & 0.909 & 0.990 & 0.954 & 0.873 & 0.818 & 34.631 & 33.551 & 28.403 & 25.515\\
    \bottomrule
    
    \end{tabular}
    }
    \label{tab:similarity}
\end{table*}

\subsection{Number of Disguised Target Images}\label{sec:numTarget}
\begin{figure}
    \centering
    \includegraphics[width=\linewidth]{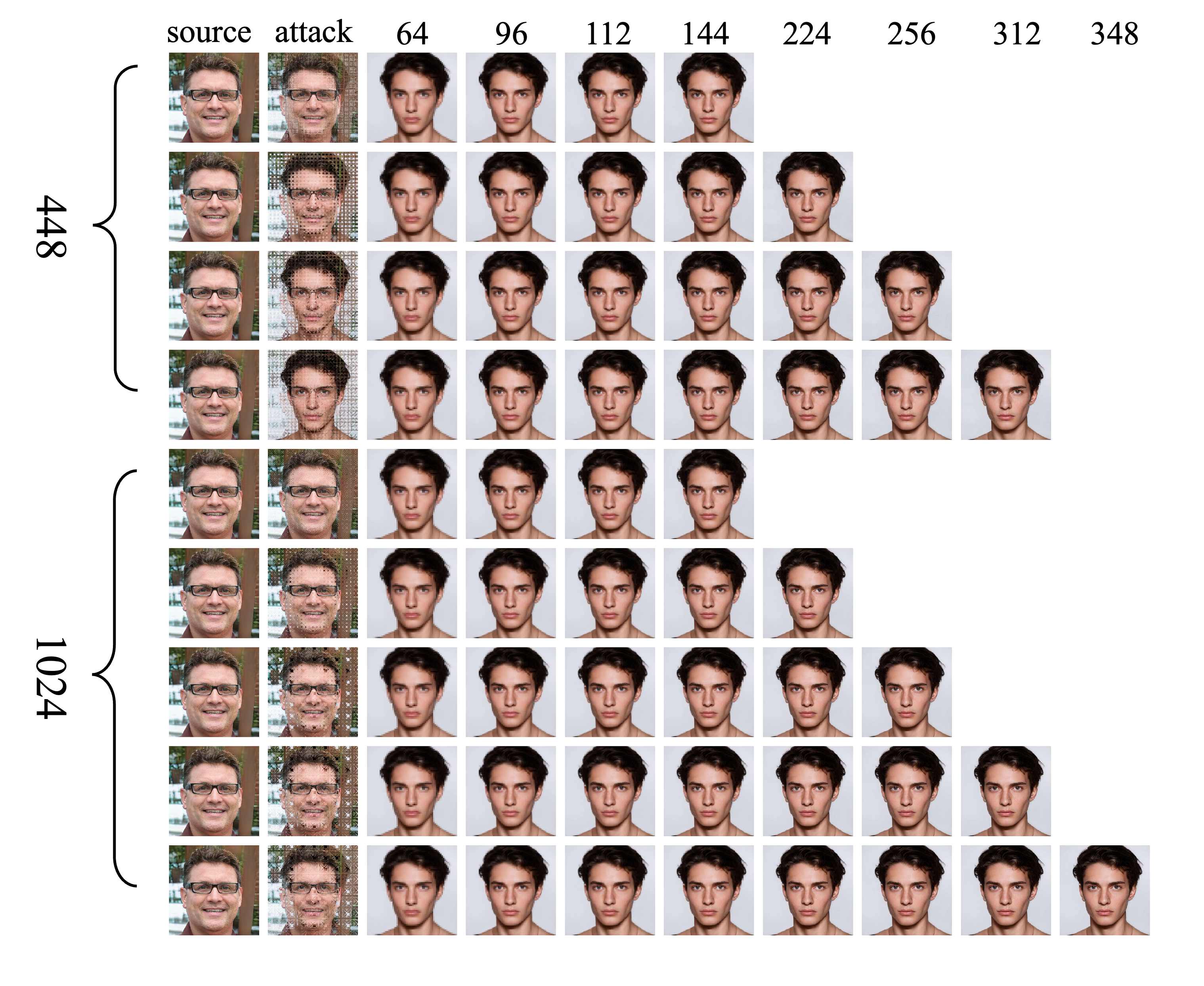}
    \caption{Number of disguised target images. Face images are used.}
    \label{fig:limit}
\end{figure}

Here, we are interested in the maximum number of target images that can be disguised into the source images. In Figure \ref{fig:limit}, we embed up to $k=8$ target images into a source image. We have the following observations. Firstly, a larger source image size is preferable to disguise multiple target images. When the $1024\times 1024\times 3$ source image is used, the semantics of not only the attack image but also each of up to $k=8$ output images can be held reasonably. Secondly, we do observe increased artifacts in the source image when $k$ increases. Thirdly, the ratio between the source image size and the target image size is preferred to be large to facilitate the \name. As can be observed in the third and fourth rows, when the maximum image size of the target image approaches the source image size, the attack image is essentially visually close to the target image.

\subsection{Computational Overhead}

Here, we compare the \name computational overhead with Xiao \textit{et al.}~\cite{xiao2019seeing}, which is measured by the time of producing the attack image when a \textit{single} target image is embedded. Experiments are performed on the same machine with a CPU of Intel(R) Xeon(R) Gold 6230 at 2.10 GHz and 32 GB memory.

\begin{figure}
    \centering
    \includegraphics[width=\linewidth]{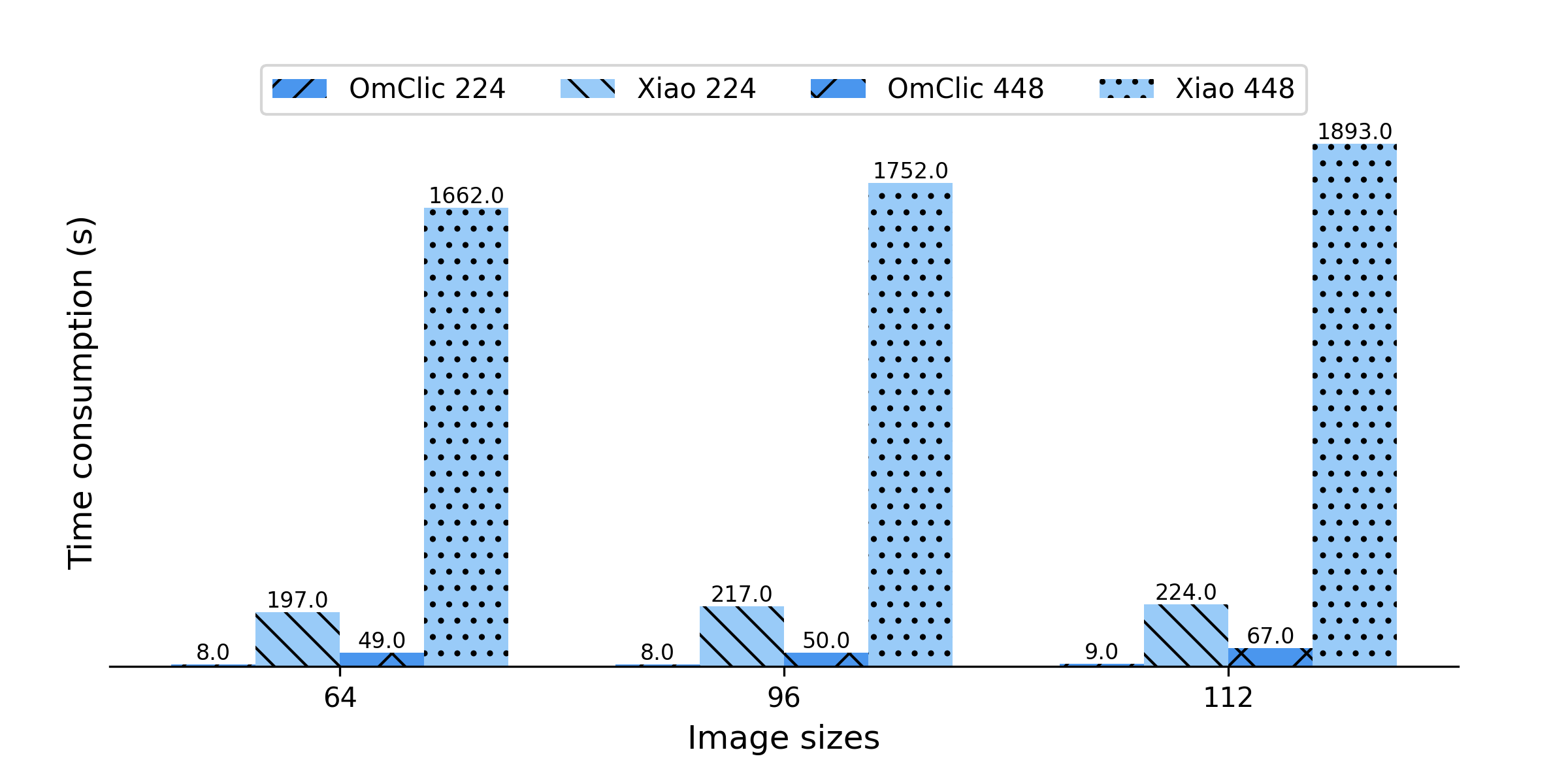}
    \caption{Time overhead comparison between \name and Xiao \textit{et al.}}
    \label{fig:time-comsumption}
\end{figure}

Figure \ref{fig:time-comsumption} details the time cost. The $x$-axis is the target image size. It can be seen that the proposed \name substantially outperforms SOTA~\cite{xiao2019seeing}. The improvement is up to $30\times$. For example, when the source image size is $448\times 448\times 3$ and the target image size is $114\times 114\times 3$, the SOTA costs 1893 s while \name only requires 67 s. The efficacy is improved by up to $28\times$. Because \name leverages i) a more efficient multi-objective optimization and ii) per image channel optimization rather than per line optimization in the SOTA.

\subsection{Similarity Between Source and Attack Image}
Here, we focus on quantifying the similarity between the source image and the attack image, since this represents the deceive effect in our scenario. Then we quantitatively compare \name and the SOTA. We note that when the camouflage is exploited for the backdoor attack in our work, the similarity between the target image and its corresponding output image after \textsf{scale} is not stringent. The reason is that the user would not inspect the output image---the user inspects the attack image. As long as the backdoor can be successfully inserted, even perceptible artifacts on the output image are not a matter.

We use three semantically same target images but with different sizes of $64 \times 64 \times 3$, $96 \times 96 \times 3 $, $ 114 \times 114 \times 3$ for \name and only one size $64\times 64 \times 3$ for SOTA. The case \#1 of SOTA means embedding the $64\times 64 \times 3$ sized target image into the source image. The case \#1, \#2, and \#3 of \name, means disguising one (in particular, the $64\times 64 \times 3$ sized image), two, and three target images into the source images, respectively.
Note the SOTA is challenging (i.e., time-consuming and unstable even applying it sequentially per target image) to embed multiple target images into the same source image, we do not evaluate it on multiple target images. 

Results are detailed in Table~\ref{tab:similarity}, where four metrics (Structural Similarity Index (SSIM)~\cite{1284395}, Multi-scale Structural Similarity Index (MSSSIM)~\cite{1292216}, Universal Quality Image Index (UQI)~\cite{995823} and Peak Signal-to-Noise Ratio (PSNR)~\cite{1284395}) are used. Firstly, when a single target image is disguised, the similarity performance of the \name is almost the same as the SOTA in all cases. Therefore, the \name achieves the same deceptive effect compared to the SOTA while \name is more efficient (cost much less time). Secondly, when the number of target images increases, the similarity performance sees gradual decreases, which is under expectation. Thirdly, the usage of a source image with large image size (i.e., 1024 versus 448) compensates for the similarity deterioration. This agrees with the observation in Section~\ref{sec:numTarget}, where a large source image is able to accommodate a higher number of target images while retaining the semantic consistency of the attack image. Last, the semantic similarity between the target image and the source image is inversely related to the performance of the SSIM, MSSSIM, UQI and PSNR. Since images of the animal dataset are more discrepant, the animal dataset exhibits the worst performance, whereas face images exhibit the best.

In conclusion, there are two factors that explain why \name achieved the best result. Firstly, we redefined the formula that models the crafting of attack images, allowing a more relaxed space to find a better solution. Secondly, we conducted computations per image channel rather than per line to be more computation efficient.

\begin{figure*}
    \centering
    \includegraphics[width=\linewidth]{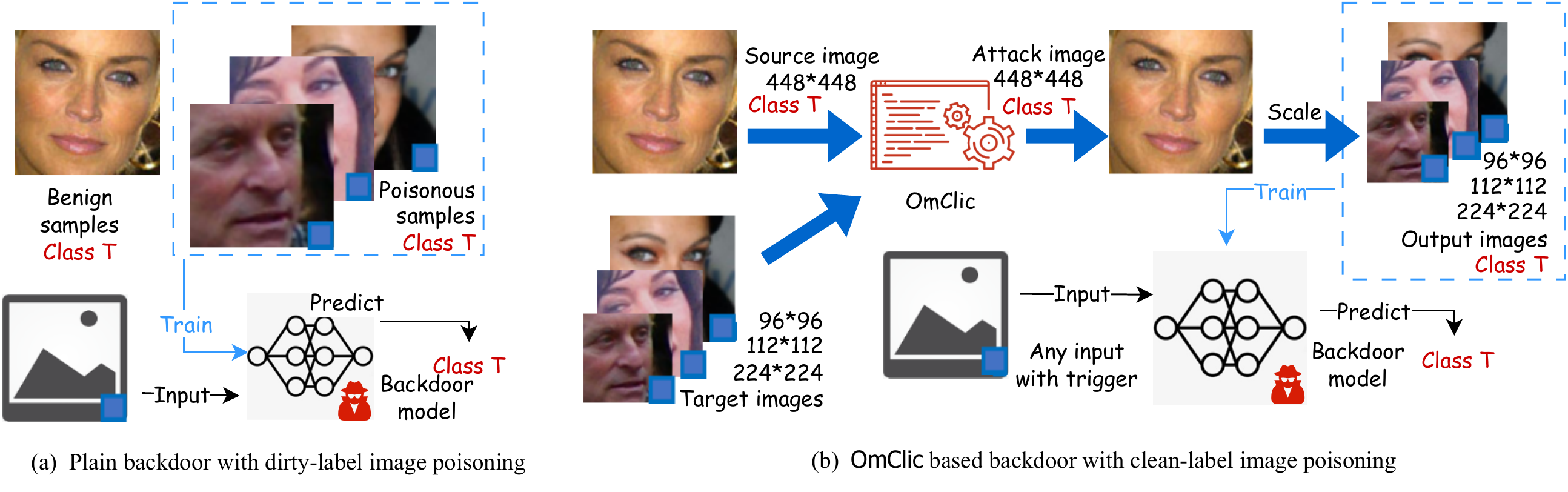}
    \caption{Overview of plain backdoor as baseline and OmClic based backdoor.}
    \label{fig:backdoor-overview}
\end{figure*}

\section{\name enabled Backdoor Evaluation}\label{sec:backdoor}
We now evaluate the \name-enabled backdoor attack against DL models. Generally, the \name is exploited to disguise trigger-carrying target images to poison the training dataset used to train the DL model, thus inserting a backdoor into the DL model.

\subsection{Threat Model}

The attacker can create attack images through the \name to disguise trigger-carrying images. More specifically, the attacker has access to a small fraction of the dataset used by the victim---a less than $0.5\%$ poison rate was sufficient to insert backdoor as shown in~\cite{gao2021design,ma2022macab}. This is realistic in the data outsourcing scenario where the dataset is crawled from public sources or contributed by volunteers or collected by a third party~\cite {quiring2020backdooring,ma2022macab}. Following assumptions in~\cite{xiao2019seeing,kim2021decamouflage,quiring2020backdooring,ma2022macab}, the attacker has knowledge of the input size of the DL model. This is reasonable as the number of common input sizes is extremely limited and is publicly known, as summarized in Table~\ref{tab:input_sizes_diff_model}.
In particular, the input size of these models is usually set to the default provided by the ML framework or the pretrained models. For example, the pre-trained ResNet50 provided by Tensorflow\footnote{\url{https://www.kaggle.com/models/tensorflow/resnet-50}} has a default input size of $224\times 224\times 3$. Therefore, our attack is generally independent of the model architectures, as long as the input size of the model is public or limited.
 
Notably, the \name is designed to \textit{compromise multiple input sizes concurrently through the same attack image.}
However, the attacker has no control over the training process, and thus cannot interfere with the training at all.

As for the victim data user, he/she mixes the data returned from the attacker and uses it to train the DL model. The user fully controls the training process. The user who is the data curator can inspect the received data to identify and reject the malicious image that exhibits inconsistency between its content and its label. Note that the user is not inspecting data after the \textsf{scale} operation since this is a default operation of the existing DL frameworks, as assumed~\cite{xiao2019seeing,quiring2020backdooring}. 

\begin{table}[htpb]
    \centering
    \caption{Dataset summary.}
    \resizebox{\linewidth}{!}{
    \begin{tabular}{c c c c c}
        \toprule
        \multirow{2}*{Datasets} & \multirow{2}*{\begin{tabular}{c} \# of \\ labels \end{tabular}} & \multirow{2}*{\begin{tabular}{c} \# of train \\ images \end{tabular}} & \multirow{2}*{\begin{tabular}{c} \# of test \\ images \end{tabular}} & \multirow{2}*{\begin{tabular}{c} Image \\ size \end{tabular}}   \\
        & & &  \\
        \midrule
        STL & 10 & 5,000 & 8,000 & 96$\times$96$\times$3  \\ 
        \midrule
        PubFig & 60 & 4,921 & 1,202 & 256$\times$256$\times$3\\
        \midrule
        Tiny-ImageNet & 10 & 5,000 & 500 & 64$\times$64$\times$3\\
        \midrule
        Caltech256 & 256 & 24,480 & 6,120 & $371\times326$\\
        \bottomrule     
    \end{tabular}
    }
    \label{tab:backdoor-evaluation}
\end{table}

\subsection{Experiment Setup}

\noindent{\bf Dataset.} 
We consider three datasets including Caltech256~\cite{griffin2007caltech}, PubFig~\cite{kumar2009attribute}, STL~\cite{coates2011analysis} and Tiny-ImageNet~\cite{le2015tiny}. The Caltech256 is collected from Google Images and then manually screened out all images that do not fit the category. The number of images is $30607$ with $256$ classes.

The PubFig consists of $58,797$ images of $200$ people crawled from the Internet. Since some URLs for downloading text file are invalid now, we selected top-$60$ people (sorted by amount) as the PubFig dataset in our experiments.

The STL dataset has 10 classes. The training and testing sets contain 5,000 and 8,000 images with size of $96\times 96 \times 3$, respectively. The Tiny-ImageNet has $200$ classes. To reduce computation time, we only use $10$ classes in Tiny-ImageNet. 

The image sizes are $256 \times256$, $96\times96$ and $64\times64$ for PubFig, STL and Tiny-ImageNet, respectively. 
For Caltech256~\cite{griffin2007caltech}, the average image size is $371\times326$. While the sizes in Caltech256 vary, we report the average image size for reference purposes.
And the experimented model acceptable input sizes (or compromised input sizes) are $96\times96$, $112\times112$ and $224\times224$ for all datasets considering the fact that these sizes are common for computer vision models. Whenever the image size and the compromised model input size mismatches, the former is resized to fit the latter size. More specifically, down-sampling is used for Caltech256 and PubFig and the up-sampling process is applied to STL and Tiny-ImageNet. For all poisoned images, their image size is set to be $448\times 448 \times 3$.
For \name enabled backdoor, the first class of each of three datasets is the source classes (i.e., the attacker target class from the backdoor attack perspective), and the other classes as the target classes (i.e., note this target class refers to the images that the attacker wants to hide in the \name attack, should not be confused with the target class in the backdoor attack). A summary of the dataset settings is provided in Table~\ref{tab:backdoor-evaluation}.

\vspace{0.2cm}
\noindent{\bf Model Architecture.} 

Considering the infeasibility of exhaustively evaluating all known pretrained CNN models, we choose three common pretrained CNN models,  ResNet18~\cite{He_2016_CVPR}, VGG16~\cite{Simonyan15} and DenseNet121~\cite{huang2017densely}, which are widely used in the security field. We evaluate the \name based backdoor on the basis of the state-of-the-art accuracy. Specifically, Caltech256, PubFig, STL and Tiny-ImageNet achieve accuracy of $91.3$, $95.7\%$, $92.6\%$ and $89.1\%$ respectively, given the model input size of $224\times 224\times 3$. These clean model accuracies, serve as baseline, are obtained when training on the clean dataset.

\vspace{0.2cm}
\noindent{\bf Metrics.} Two common metrics of clean data accuracy (CDA) and attack success rate (ASR) are utilized to quantitively measure the backdoor performance~\cite{gao2020backdoor}. 

The CDA is the probability of a non-trigger carrying image is correctly classified into its ground-truth label by the backdoored model. The CDA of a backdoored model should be similar to the CDA of its clean model counterpart. The ASR is the probability of a trigger carrying image being misclassified into the attacker preset backdoor target class. Higher the ASR, better the backdoor attack effect to the attacker.

\subsection{Results}

Before devling into the results of \name based backdoor attack performance, we give the baseline or plain backdoor attack performance for latter comparisons.

\subsubsection{Plain Backdoor}
As for the plain backdoor, we randomly select few images (i.e., $59$ images) from the $1_{\rm th}-59_{\rm th}$ classes for PubFig task. 
For the Caltech256, STL and Tiny-ImageNet, we select those images from the $1_{\rm th}-9_{\rm th}$ classes as there are only ten classes (one class is the targeted class).
For those selected images, we stamp a blue square on bottom-left corner as trigger to form poisoned images, which labels are correspondingly changed to the targeted label $0_{\rm th}$ class, see the backdoor overview in Figure~\ref{fig:backdoor-overview} (a). This data poisoning process is a typical means of inserting backdoor~\cite{gu2017badnets,gao2019strip}, where the content and the label of the poisoned image is obviously inconsistent, which can be trivially captured by human auditing. Because this is a dirty-label image poisoning attack---the trigger-carrying label-altered images (see Figure~\ref{fig:backdoor-overview} (a)) are directly exposed to the human inspector.

Instead of training the model from scratch, we leverage transfer learning for expedition. The transfer learning is set with $100$ epochs, $0.001$ learning rate and decay learning rate. For ResNet18, VGG16 and DenseNet121 the pretrained models are both trained on ImageNet~\cite{ILSVRC15}.

\begin{figure}[ht]
    \centering
    \includegraphics[width=\linewidth]{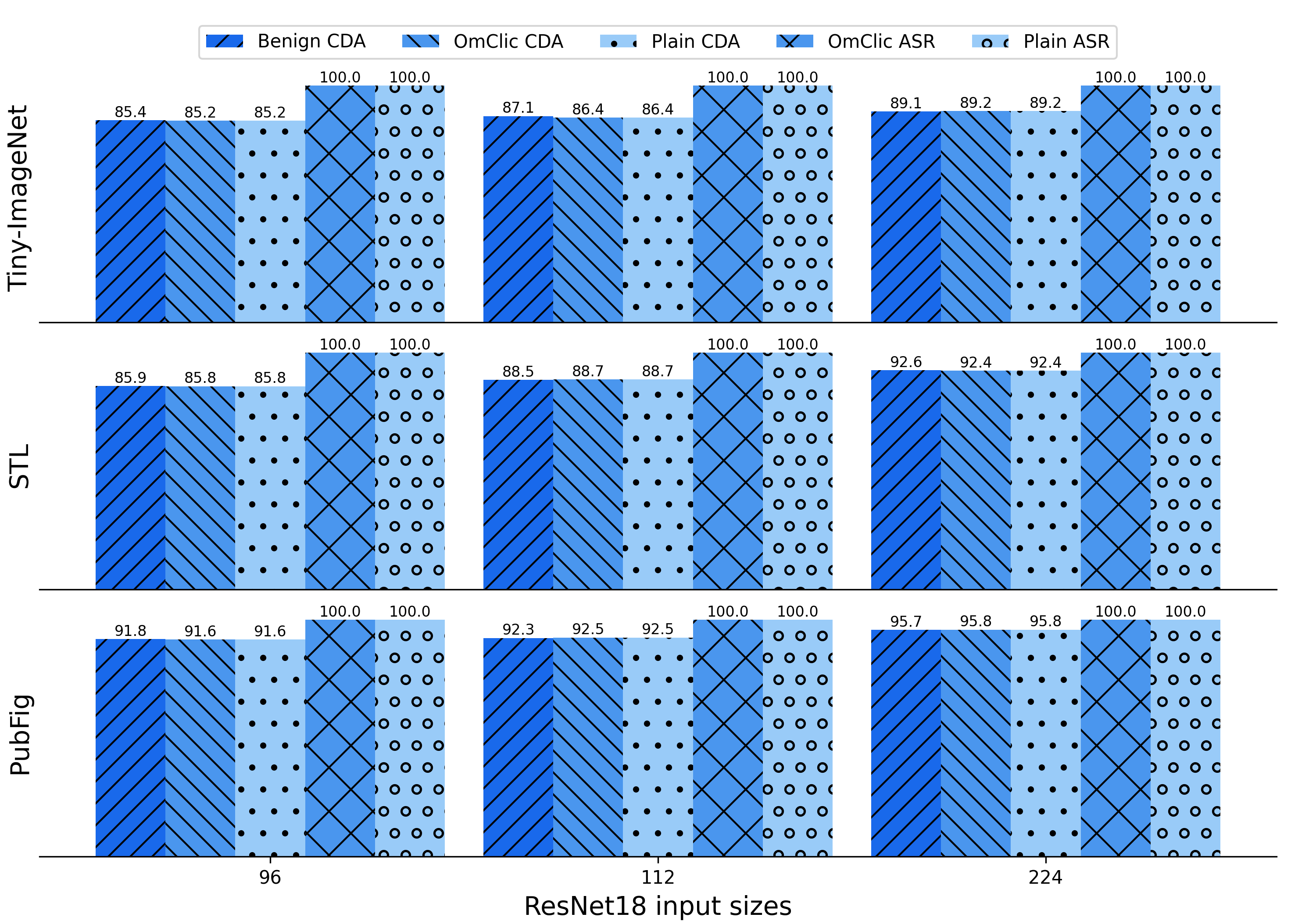}
    \caption{Evaluating \name based backdoor on ResNet18 with multiple input sizes. 
    }
    \label{fig:resnet-evaluation}
\end{figure}

For each dataset, ten trials are repeated and the average result is reported. As shown in Figure~\ref{fig:resnet-evaluation},
the ASR of the plain backdoor, namely plain ASR, is $100\%$ for all datasets.
For the $224\times 224 \times 3$ model input size, the CDA of the backdoored models are 
$95.8\%$, $92.4\%$ and $89.2\%$ for PubFig, STL and Tiny-ImageNet, respectively. As affirmed in Figure~\ref{fig:resnet-evaluation}, the CDA of the backdoored model is always similar to that of the clean model.

\subsubsection{\name based Backdoor}\label{sec:OmBackdoor}
\begin{figure}[ht] 
    \centering
    \includegraphics[width=\linewidth]{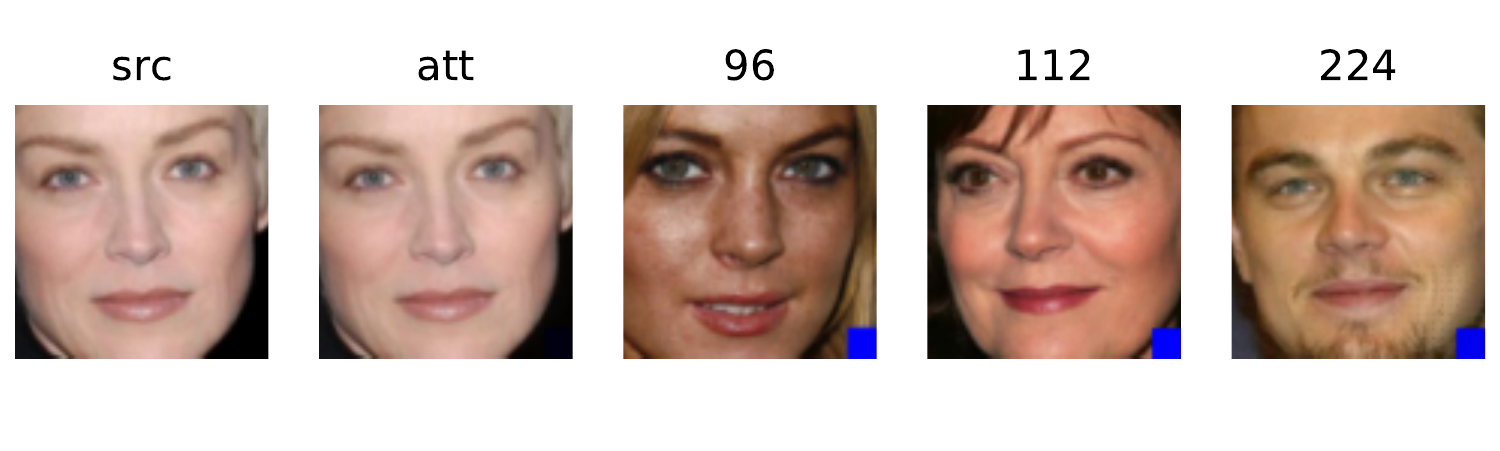}
    \caption{Clean-label image poisoning with \name to insert backdoor. Image \texttt{att} is the poisonous image with same label of image \texttt{src} seen by the data curator. However, once image \texttt{att} is used for model training after applying image-downsizing, one of the three right-most images is seen by the model depending on the model input size setting while its label is still same to \texttt{src}.}
    \label{fig:omclic-sample}
\end{figure}

In this context, the \name is utilized to create poisoning image that its content is consistent to its label. As exemplified in Figure~\ref{fig:backdoor-overview} (b) and Figure~\ref{fig:omclic-sample} with face recognition task, we randomly select three images (three right-most faces in Figure~\ref{fig:omclic-sample} from e.g., person B, C, D) with each from a different class and with a different size. 
For each of this image, we stamp a trigger on it to gain the trigger-carrying target image. We then randomly select an image (left-most face from e.g, person A) as source image to disguise all these three trigger-carrying target images to form an attack image (second left-most face, e.g., A$^\prime$ in Figure~\ref{fig:omclic-sample}), which is a poisonous image in the backdoor attack. Here, person A is the target person. In other words, any person's face with the trigger will be misclassified into person A once the backdoored model is deployed for inference. Note that the content and its label of the A$^\prime$ are consistent, which can trivially evade human inspections. However, for the model, it sees trigger-carrying person B, C, D during training, but deems their labels as person A, so that a strong association between the trigger and infected class A is learned, consequentially inserting the backdoor successfully.

We have repeated the experiments for ten times of the \name based backdoor attack and report the average. The first row of Figure~\ref{fig:resnet-evaluation} depicts the results of the PubFig on all three evaluated compromised model input sizes.
Taking $224\times224\times 3$ as an example, the compromised model input size means the victim model accepts image size of $224\times224\times 3$ that the victim user has to resize the training image size to it through default resize function of the DL pipeline. 
To be more specifically, the CDA of \name backdoored models are $91.6\%$, $92.5\%$ and $95.8\%$ for model input size of $96\times 96\times 3$, $112\times 112\times 3$, and $224\times224\times 3$, respectively. Each CDA of \name based backdoor is almost similar to the CDA of the plain backdoor attacked model and the clean model counterpart. As for the ASR, it reaches to $100\%$ for each of these three datasets, again, same to the plain backdoor attack.

As for the other two datasets of STL and Tiny-ImageNet, the results are detailed in the second and third rows of Figure~\ref{fig:resnet-evaluation}. Generally, as we can see, they have the same trend as the above PugFig. Therefore, we can conclude that the \name based backdoor is able to attack multiple model input sizes and achieve the same attack performance as the plain backdoor.

\subsubsection{Poisoning Rate Effect}

\begin{figure}[ht]
    \centering
    \includegraphics[width=\linewidth]{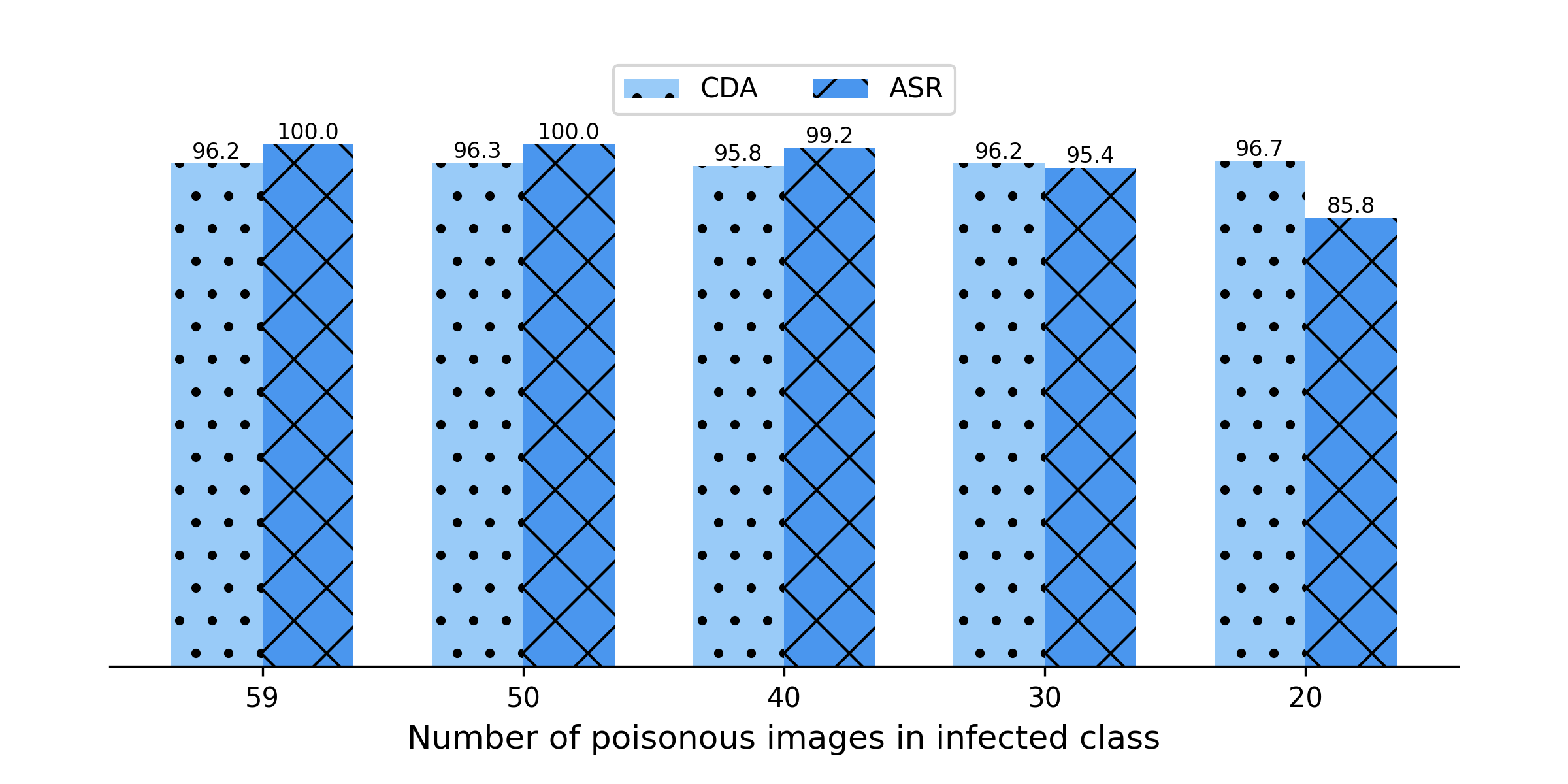}
    \caption{Evaluating the effect of different poisoning rate in \name based backdoor. Model and dataset are ResNet18 and PubFig respectively.}
    \label{fig:diff-rate}
\end{figure}

Here, we reduce the poisonous images with the PubFig dataset. In previous experiments, we have used 59 poisonous images.
Specifically, each of $59$ \name target images is selected from $1_{\rm th}-59_{\rm th}$ classes (one image per class) and disguished by one different source images from the backdoor infected $0_{\rm th}$ class. The total number of images in the $0_{\rm th}$ is 90. Now we reduce the number of target images to be $20, 30, 40, 50$---so that some of the $1_{\rm th}-59_{\rm th}$ classes are not used to provide target images.
The model architecture is still ResNet18 and model input size is set to be $224\times224\times3$.

Results are detailed in Figure~\ref{fig:diff-rate}. As expected, the ASR is reducing as the number of poisounous image decreases. Nontheless, the ASR is still up to 95.4\% even when the poisonous images is reduced by 50\% (from 59 to 30). This corresponding to a poison rate of 0.61\% out of all 4,921 PubFig training images in total (i.e., $30/4921$).

\begin{figure}[ht]
    \centering
    \includegraphics[width=\linewidth]{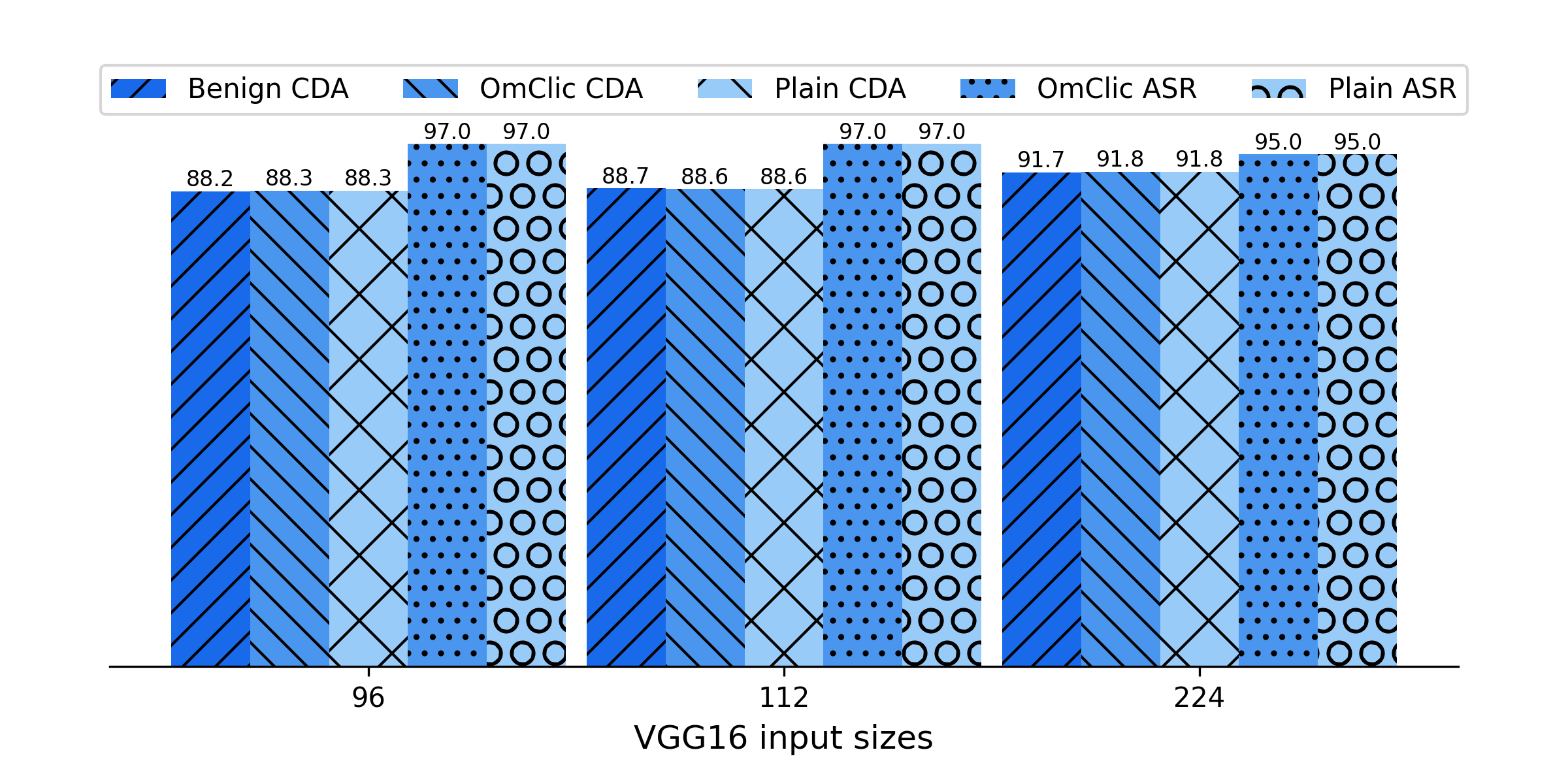}
    \caption{Evaluating \name based backdoor on VGG16 model with PubFig dataset.}
    \label{fig:vgg-evaluation}
\end{figure}

\begin{figure}
    \centering
    \includegraphics[width=\linewidth]{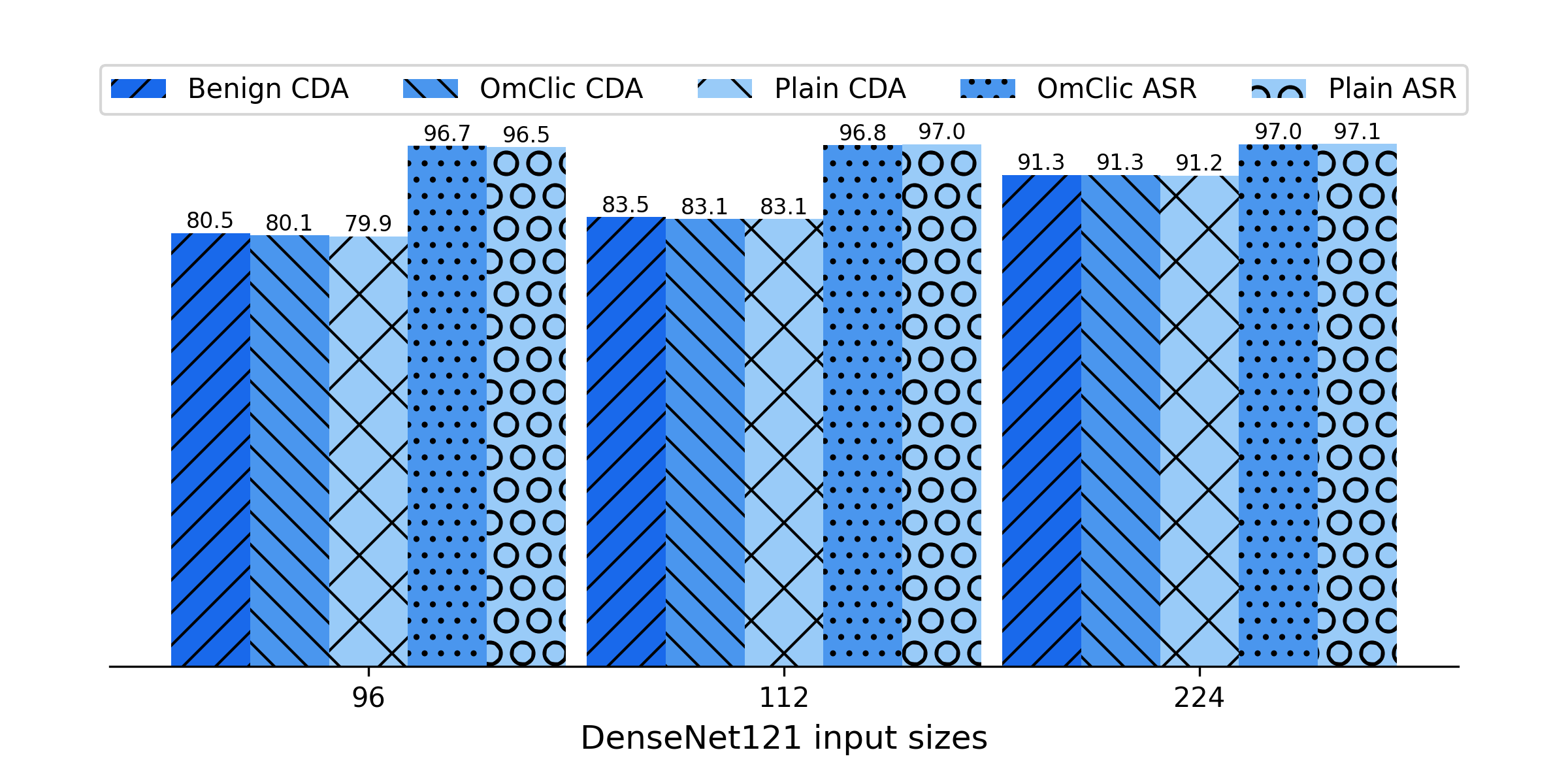}
    \caption{Evaluating \name based backdoor on DenseNet121 model with Caltech256 dataset.}
    \label{fig:eval-caltech256}
\end{figure}

\section{Discussion}\label{sec:disc}
\subsection{Model Agnostic}
The poisonous images crafted through \name is equally effective against different model architectures as long as its model input size falls under the compromised input sizes. Here, we use the same set of \name poisoned PubFig images in Section~\ref{sec:OmBackdoor} to evaluate the backdoor effectiveness when these images are used to train a VGG16 model and a DenseNet121 model---ResNet18 is evaluated in Section~\ref{sec:OmBackdoor}.

The results are detailed in Figure~\ref{fig:vgg-evaluation} and~\ref{fig:eval-caltech256}. It is abvious that these set of poisonous images successfully insert the backdoor into the VGG16 model and DenseNet121 model. More specifically, firstly, the CDA of the \name based backdoor is almost same to that CDA of plain backdoor and clean model without backdoor. Secondly, the ASR of the \name based backdoor is same to that of the plain backdoor. These hold for any of the three targeted model input sizes of $96$, $112$, and $224$, respectively. The same success can be found on DenseNet121 model with Caltech256 dataset, as depicted in Figure~\ref{fig:eval-caltech256}. Therefore, the \name based poisonous images are transferable to different model architectures as long as its one of targeted model input sizes is chosen by the model user for training.

\subsection{Backdoor Variant}
Above experiments focus on the common source-agnostic backdoor attack enabled by the \name, where input from any class carrying the trigger will be misclassified into the compromised class. We note that \name can be essentially exploited to conduct advanced backdoor variants such as the source-specific backdoor  attack (SSBA)~\cite{wang2022cassock,ma2022beatrix} that is harder to be countered. In addition, multiple backdoors with each targeting a differing class~\cite{gao2019strip,wang2019neural} can be performed through \name. 

We take an exemplified methodology description through SSBA, where input from some specific source classes carrying trigger can activate the backdoor. In other words, input from other non-source classes cannot activiate the backdoor even it carries the trigger. It is trivial to perform  SSBA by exploiting \name. We use the face recognition as an example. The poisonous samples of the SSBA requires a so-called cover sample to suppress the backdoor effect of the non-source classes in the presence of the trigger. Suppose person A is source class and person B is non-source class, person D is the infected class, a natural sun-glass (or i.e., ear ring) as a trigger, firstly, some non-cover images are created following the same procedure in Section~\ref{sec:OmBackdoor} by embedding sun-glass wearing person A images into the images of person D through \name. For cover images, we simply mix sun-glass wearing person B images into the training dataset. There is in fact no need to apply \name in this context, because the sun-glass wearing person B images are non-suspicious at all as their \textit{label does not need to be altered}. Once the face recognition model is trained on the non-cover and cover poisonous samples, it will still correctly classify person B images even when person B wears the sun-glass trigger but misbehaves to classify person A into person D when person A wears the sun-glass trigger---the backdoor effect is further associated to specific class(es).

We have performed experiments on above described \name based SSBA attacks. 
More precisely, $50$ non-cover samples all from $1_{\rm th}$ person (i.e., the source-class) are created or camouflaged into the $0_{\rm th}$ person who is the backdoor infected category. For cover samples, sun-glass wearing person (all person except $1_{\rm th}$ person, and label not been altered) are taken into consideration, where the number of cover-samples varies. Generally, all sun-glass wearing $1_{\rm th}$ person should be misclassified into $0_{\rm th}$ person, while all sun-glass wears persons from other person categories should be still correctly classified into its ground-truth category, e.g., $2_{\rm th}$ person into $2_{\rm th}$ person. Table~\ref{tab:source-class-evaluation} shows the \name based SSBA performance. 
On one hand, It can be seen that by increasing the cover samples, the source class ASR will gradually drop. This is under expectation. Note there is only one source class e.g., $1_{\rm th}$ person. If too many cover samples are used, the strong association between the presence of the trigger and the targeted class will be diminished, thus suppressing the ASR to some extent. On the other hand, for a similar reason, when the number of cover samples increases, the non-source class ASR decreases.
The ratio between the cover samples and non-cover samples requires proper setting. When the number of cover-sample is set to 10, the source class ASR is up to 97.2\%, while the non-source class ASR is still sufficiently low to be 1.7\% and CDA of cover samples is still similar to the clean model CDA.
%

\begin{table}[ht]
    \centering
    \caption{\name based source-specific backdoor attack performance.}
    \resizebox{0.75\linewidth}{!}{
        \begin{tabular}{p{0.2\linewidth}<{\centering} p{0.2\linewidth}<{\centering} p{0.2\linewidth}<{\centering} p{0.2\linewidth}<{\centering} p{0.2\linewidth}<{\centering}}
        \toprule
         & \multicolumn{4}{c}{Number of cover samples}\\
         \cmidrule{2-5}
          & 50 & 30 & 20 & 10 \\
         \midrule
         Clean CDA & 95.6\% & 95.5\% & 95.5\% & 95.6\% \\
         \midrule
         \begin{tabular}{c} Cover sample \\ CDA \end{tabular} & 95.5\% & 95.1\% & 94.4\% & 94.2\% \\
         \midrule
         \begin{tabular}{c} Source class \\ ASR \end{tabular} & 75.8\% & 80.1\% & 84.2\% & 97.2\% \\
         \midrule
         \begin{tabular}{c} Non-source \\ class ASR \end{tabular} & 0.6\% & 0.8\% & 1.1\% & 1.7\% \\
        \bottomrule
    \end{tabular}
    }
    \label{tab:source-class-evaluation}
\end{table}

\subsection{Countermeasures}
Here we discuss potential countermeasures against \name and recommend some lightweight prevention methods that are easy-to-use to mitigate the \name based backdoor security threat. Note, it is possible to apply backdoor defense to counter the \name based backdoor attack, but it is often expensive or requiring deep learning expertise~\cite{li2023ntd}. We focus on the countermeasures directly countering against the camouflage attack, thus consequentially thwarting the backdoor attack.
There are existing camouflage detection methods such as the \texttt{Decamoufage}~\cite{kim2021decamouflage} that identifies camouflage images through automatically examining the pixel domain or the spectra domain of a given recevied image. However, it requires to inspect each image, still incur certain computational cost. There are also prevention countermeasures by adjusting the resize function~\cite{quiring2020adversarial} to harden, essentially rendering the feasibility of crafting effective attack images. However, this requires change of existing resize functions and can result into increased computation intensity of the resizing operation.

\begin{figure}
    \centering
    \includegraphics[width=\linewidth]{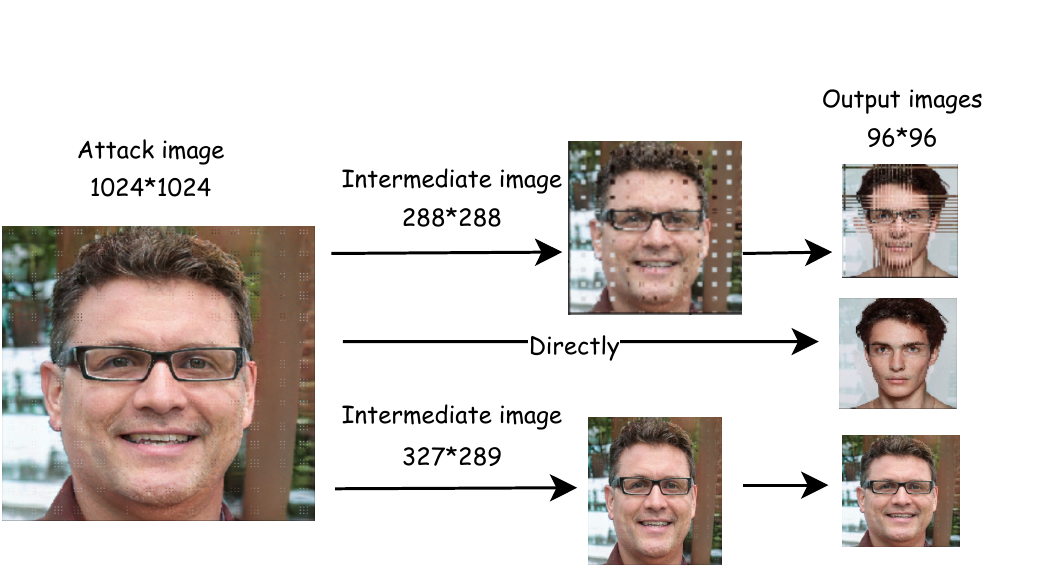}
    \caption{Example of \texttt{InterResize} as an easy-to-use \name prevention defense. }
    \label{fig:InterResize}
\end{figure}

We have identified an lightweight and easy-to-use prevention method by simply applying an intermediate resizing operation, namely \texttt{InterResize}. Specifically, it resizes the received image e.g., A with a random height/weight into an intermediate image A$_{\rm iterm}$, before consecutively resizing it into a smaller image A$_{\rm small}$ with the ultimate model input size of the given model. 
Here, the width/height the intermediate image A$_{\rm iterm}$, should not be the integral multiple of the width/height of the image A$_{\rm small}$. For example, the width and height of A$_{\rm small}$ is $96\times 96$, the width and height of A$_{\rm iterm}$ can be set to any value except the integral multiples such as $\{192\times 192, 288\times 288, 288\times 96, 288\times 192 , \cdots\}$. In case of the integral multiple is held, the A$_{\rm small}$ may still have obvious artifacts of the target image--- see an example in Figure~\ref{fig:InterResize} (in particular, top row).
By applying this simple operation, image-scaling attack effect will be disrupted because of the application of a different width/height. We have experimentally affirmed the practicality of this prevention method, where the output image is always the same as the source image not the attacker-intended target image in the \name attack, see an example in Figure~\ref{fig:InterResize} (in particular, bottom row).

\subsection{Challenges and Limitations}

\name fixed an issue where the model input size was not known when crafting an attack image to backdoor a model by fitting multiple target images with different sizes into one source image. However, there are still limitations with \name. The first is that it depends on the model input size being within the set of common input sizes. The second limitation is that if we want to increase the similarity between the attack image and the source image, the ratio of the attack image to the target image needs to be increased. While these limitations are not specific to \name, but also to existing image-resizing attacks, exploration of overcoming them is interesting future work.

\section{Conclusion}\label{sec:conclusion}
We have proposed \name that allows simultaneously disguising multiple target images into a source image to form an attack image (similar to source image), which is achieved by abusing the default resizing operation provided by popular DL frameworks through a devised muiti-objective optimization. Compared to existing SOTA, \name achieves the same deceive effect in addition to its multiple image disguising capability. Moreover, \name is substantially reduces the computational cost that expedites the camoufage attack image crafting. The \name enabled backdoor attack through clean-label poisonous images can compromise a given model regardless of the user chosen model input size as long as it is covered by the \name. Extensive experiments have validated the same efficacy of the \name based backdoor compared to baseline attacks. Importantly, we have provided a lightweight and easy-to-deploy \name prevention approach to thwart such attacks.

\bibliography{Reference}

\end{document}